\documentclass[11pt]{article}
\usepackage[utf8]{inputenc}
\usepackage[left=2.54cm, right=2.54cm, top=2.54cm, bottom = 2.54cm]{geometry}
\usepackage{xcolor} 
\usepackage{indentfirst} 
\usepackage{caption}
\usepackage{hyperref}
\usepackage{mathtools}

\usepackage{mathrsfs}
\usepackage{amsmath, amssymb, amsfonts, amsthm,rotating,booktabs,array}
\usepackage{cancel}
\usepackage{bm}
\usepackage{booktabs}
\usepackage{subfigure}
\usepackage{graphicx}
\usepackage{mathabx}
\usepackage{setspace}
\usepackage{overpic}

\usepackage{algorithm,algpseudocode}
\algnewcommand\algorithmicinput{\textbf{Input:}}
\algnewcommand\algorithmicoutput{\textbf{Output:}}
\algnewcommand\Input{\item[\algorithmicinput]}%
\algnewcommand\Output{\item[\algorithmicoutput]}%

\usepackage{xr}
\makeatletter
\newcommand*{\addFileDependency}[1]{
\typeout{(#1)}
\@addtofilelist{#1}
\IfFileExists{#1}{}{\typeout{No file #1.}}
}\makeatother

\newcommand*{\myexternaldocument}[1]{%
\externaldocument{#1}%
\addFileDependency{#1.tex}%
\addFileDependency{#1.aux}%
}

\myexternaldocument{GNet_arXiv_SI}

\makeatletter
\def\algbackskip{\hskip-\ALG@thistlm}
\makeatother
\usepackage{mathtools}
\usepackage{multicol}
\usepackage{multirow}
\usepackage{verbatim}
\usepackage{enumerate}
\usepackage{relsize}
\usepackage{listings} 

\newtheorem{theorem}{Theorem}[section]
\newtheorem{corollary}{Corollary}[section]
\newtheorem{lemma}{Lemma}[section]

\newtheorem{assumption}{Assumption}[section]
\newtheorem{example}{Example}[section]
\theoremstyle{plain}

\usepackage{ifthen}

\newcommand{\mbold}{1}

\ifthenelse{\mbold=1}{ 
\newcommand{\mathbold}{\mathbf} 
\newcommand{\bmbold}{\bm}}
{\newcommand{\mathbold}{} 
\newcommand{\bmbold}{}
}

\title{GNet: A scalable and flexible Gaussian process network with nonparametric neurons}
\author{Mengyang Gu\footnote{Correspondence should be addressed to Mengyang Gu (\href{mailto:mengyang@pstat.ucsb.edu}{mengyang@pstat.ucsb.edu})}\\ \\
      \small   Department of Statistics and Applied Probability, University of California, Santa Barbara, CA }
\date{}

\begin{document}

\maketitle

\begin{abstract}
We build GNet, a scalable and flexible Gaussian process network with nonparametric activation functions. The key computational contribution is to make estimation of GP activations computationally scalable  without forming covariance matrices explicitly. We provide theoretical studies comparing function classes with a fixed architecture, and derive  predictive risk bounds for GNets. Using essentially the same  architecture and optimization settings,  GNet shows competitive performance across a diverse range of test problems, including predicting nonlinear functions, nonparametric regression of real-world data, and predicting one-body direct correlation functions with high-dimensional inputs in classical density functional theory.  The strong performance of  GNet suggests that flexible nonparametric neurons can substantially reduce the number of trainable parameters while remaining computationally scalable.    
\end{abstract}

\section{Introduction} 
\label{sec:intro}

Modern predictive models often gain flexibility by increasing training data, model size, and computation. This strategy has enabled highly capable neural networks, including transformer architectures \cite{vaswani2017attention}, but the cost of training such models can exceed the resources of academic research teams, and it also remains a fundamental bottleneck in industrial applications. This raises a basic question: can some  flexibility usually supplied by more neurons and layers instead be placed within individual neurons? We address this question by making  the Gaussian process network (GNet) with nonparametric activations computationally scalable. 

Gaussian processes (GPs) and neural networks offer complementary approaches to nonlinear prediction. GPs provide efficient prediction with uncertainty quantification while requiring relatively few numerically optimized parameters.  They are commonly used for  nonparametric regression  \cite{rasmussen2006gaussian}, probabilistic integration \cite{briol2019probabilistic}, computer model emulation and calibration \cite{santner2003design,kennedy2001bayesian,sung2024review}. However, GPs have two limitations. First, storing an  $n\times n$ covariance matrix costs $\mathcal O(n^2)$, and direct factorization for matrix inversion costs $\mathcal O(n^3)$.  Second, prediction is governed by a prespecified distance metric between inputs, which can be restrictive for complex, high-dimensional inputs. 
Scalable GP approximation approaches were developed to reduce the computational cost \cite{vecchia1988estimation,snelson2005sparse,lindgren2011explicit,datta2016hierarchical,katzfuss2021general,zhu2024radial}, while deep GPs increase flexibility through hierarchical models of covariance functions \cite{damianou2013deep,sauer2023vecchia}. Both directions involve  computation-accuracy tradeoffs as data size or latent complexity increases.

Neural networks instead build flexibility through compositional structures. A standard neuron forms a one-dimensional (1D) projection and applies a fixed activation function, such as rectified linear unit (ReLU), softplus or Gaussian Error Linear Units (GELU) \cite{hendrycks2016gelu}. With a fixed activation function, additional complexity is typically obtained with more neurons, layers, and parameters, increasing both estimation and computational costs.  GP-modeled activation functions were proposed in \cite{urban2017gaussian,manzhos2023neural} for 
learning activation functions \cite{agostinelli2014learning}.   
These approaches, however, do not exploit  state-space representation of GP activation to accelerate computation. 
Fast linear algebra for ordered state-space kernels, including those for half-integer Mat{\'e}rn kernels was studied \cite{chandrasekaran2005some,hartikainen2010kalman,nickisch2018state}, while these studies primarily focus on one-dimensional ordered inputs. GNet is also related to  generalized additive models \cite{hastie1986generalized} and additive Gaussian process regression \cite{yang2015minimax,delbridge2020randomly}. These methods also do not address scalable learning of GP-modeled  activations. The key observation is that each activation receives 1D projected inputs, enabling covariance matrix--vector products and their derivatives to be scalably computed without approximation, despite potentially high-dimensional  input variables.

Our first contribution is a scalable computational framework for learning GP activations. We  model  activations with GPs whose kernels admit state-space representations and develop the \emph{jointly inverse Kalman filter} (JIKF) algorithm, which multiplies an $n\times n$ state-space covariance matrix by a vector in $\mathcal O(nq^2)$ operations after input sorting, where $q$ is a small latent state dimension, versus $\mathcal O(n^2)$ in direct computation. For the exponential kernel and Mat{\'e}rn kernel with smoothness parameter $2.5$, $q=1$ and $q=3$, respectively. Crucially, JIKF can be applied to individual neurons whose 1D projected inputs induce different  orderings. Furthermore, we derive closed-form  gradients of the parameters, enabling JIKF to be integrated into Nystr{\"o}m-preconditioned conjugate gradient solvers \cite{rudi2017falkon} for covariance matrix-free training and prediction,  without retaining large computational graphs from automatic differentiation \cite{baydin2018automatic}. After one-time preprocessing of a trained model, the predictive-mean cost for each new input depends only logarithmically on $n$. 

Our second contribution studies how flexible activations shift model complexity from network size to  neurons themselves. We show that, under standard conditions, a GNet can approximate any feedforward neural network with conventional activations under a fixed architecture.   We construct a function representable by the GNet, but it cannot be approximated arbitrarily well by the neural networks with ReLU activation and the corresponding fixed architecture. Furthermore, we derive predictive risk bounds for GNet. Together, these results show that model complexity can come not only from adding neurons and layers, but also from learning richer functions within neurons.

Numerical studies connect these computational and theoretical results to strong predictive performance. GNet is motivated in part by predicting the one-body direct correlation function in classical density functional theory \cite{sammuller2023neural}, where the inputs are high-dimensional. We compare GNets with feedforward neural networks, Kolmogorov--Arnold networks \cite{liu2025kan},  exact GPs \cite{gu2018robustgasp}, deep GPs  \cite{damianou2013deep},  Vecchia and scaled-Vecchia GP approximations \cite{katzfuss2021general,katzfuss2022scaled} across simulated functions, real-world measurements \cite{kelly2023uci}, and the density functional application. These studies examine both predictive accuracy and computational scaling over a wide range of sample sizes and input dimensions.

The rest of the article is organized as follows. Section \ref{sec:GNet} introduces the GNet and JIKF.  Section \ref{sec:model_training} develops closed-form gradients and accelerated model training. Section \ref{sec:theory} studies function classes and predictive risk bounds. Section \ref{sec:numerical} presents numerical studies and  Section \ref{sec:conclusion} concludes with future directions. Derivations of all lemmas and theorems  are provided in the Supplementary Material.

\section{Gaussian process networks accelerated by a fast algorithm}
\label{sec:GNet}
A Gaussian process network (GNet) replaces parametric activation  with nonparametric activation modeled by Gaussian processes (GPs).  The output $y(\mathbold x)\in \mathbb R$ at any input $\mathbold x \in \mathbb R^p$ 
follows 
\begin{align}
y(\mathbold x)&= \sum^{d_L}_{j=1} \sigma_j  z_{L,j}(\mathbold x)+ \epsilon, 
\label{equ:gnet}
\end{align}
where $\epsilon \sim \mathcal{N}(0,\sigma^2_0)$ is an independent Gaussian noise, 
$\sigma_j$ is  a scale parameter in the final layer. Here the vector $\mathbold z_l(\mathbold x)=(z_{l,1}({\tilde x}_{l,1}),...,z_{l,d_l}({\tilde x}_{l,d_l}))^T$ contains   $d_l$  activation values at layer $l$, where the 1D input to neuron $j$ is ${\tilde x}_{l,j}=\mathbold w^T_{l,j} \mathbold z_{l-1}(\mathbold x)$, with a weight vector $\mathbold w_{l,j} \in \mathbb R^{d_{l-1}}$, for neuron $j=1,...,d_l$ and layer $l=1,...,L$. Activation functions are independently modeled by  zero-mean, unit-variance GPs, $z_{l,j}(\cdot) \sim \mathrm{GP}(0,\, c_{l,j}(\cdot,\cdot))$,  with   kernel  $c_{l,j}(\cdot,\cdot)$, and their scales are represented by the network parameters.  Furthermore, let the initial activation be the  input, $\mathbold z_{0}(\mathbold x)=\mathbold x$, and $d_0=p$. Model  (\ref{equ:gnet}) can be extended to include a mean function or  heteroscedastic noise, which are not studied herein.

We assume each kernel $c_{l,j}$  admits a linear state-space model or a dynamic linear model (DLM) \cite{West1997,durbin2012time}. This includes   Mat{\'e}rn kernels with  half-integer smoothness $\nu=(2q-1)/2$ with $q\in \mathbb N^{+}$ \cite{hartikainen2010kalman}. 
 For instance, a GP having the  Mat{\'e}rn kernel with     $\nu=2.5$ at any  $\tilde x_{l,j}$ and $\tilde x'_{l,j}$ follows 
\begin{align}
c_{l,j}(\tilde x_{l,j}, \tilde x'_{l,j})=\left(1+{{5}^{\frac{1}{2}}\beta_{l,j}|\tilde x_{l,j}- \tilde x'_{l,j}| }+\frac{5\beta_{l,j}^2|\tilde x_{l,j}- \tilde x'_{l,j}|^2}{3}\right)\exp\left(-{{5}^{\frac{1}{2}} \beta_{l,j}|\tilde x_{l,j}-\tilde x'_{l,j}|}\right) \,, 
\label{equ:matern_5_2}
\end{align}
where $\beta_{l,j}$ is an inverse range  parameter, for $j=1,...,d_l$ and $l=1,...,L$. The Mat{\'e}rn kernel in (\ref{equ:matern_5_2}) has been used as a default kernel class in software packages of GP predictive models \cite{roustant2012dicekriging,gu2018robustgasp}. When the smoothness parameter is $\nu=1/2$,  the  Mat{\'e}rn kernel becomes an exponential kernel: 
\begin{align}
c_{l,j}(\tilde x_{l,j}, \tilde x'_{l,j})=\exp\left(-{\beta_{l,j}|\tilde x_{l,j}-\tilde x'_{l,j}|}\right). \,
\label{equ:exp}
\end{align}

Assume that we have $n$ observations, $\mathbold y=(y_1, ...,y_n)^T$, with $y_i=y(\mathbold x_i)$  for $i=1,...,n$. Integrating out the independent GPs for the final-layer neuron activations, $\mathbold z_L$, and conditional on the final-layer 
inputs $ \tilde{\mathbold x}_{L}=(\tilde{\mathbold x}_{L,1},...,\tilde{\mathbold x}_{L,d_L})$ with $\tilde{\mathbold x}_{L,j}=({\tilde x}_{L,j,1},...,{\tilde x}_{L,j,n})^T$, for $j=1,...,d_L$, inverse range parameters $\bmbold \beta_{L,1:d_L}=(\beta_{L,1},...,\beta_{L,d_L})^T$ and the scale parameters $\bmbold \sigma^2_{0:d_L}=(\sigma^2_0,\sigma^2_1,...,\sigma^2_{d_L}) $,  the   distribution of the observations follows a multivariate normal distribution
\begin{align}
\mathbold y\mid \tilde{\mathbold x}_{L}, \bmbold  \beta_{L,1:d_L}, \bmbold \sigma^2_{0:d_L} \sim \mathcal{N}\left( \mathbold 0_n, \bmbold{\tilde \Sigma}_L \right), 
\label{equ:gnet_lik}
\end{align}
where the covariance  follows an additive form,  $\bmbold{\tilde \Sigma}_L= \sum^{d_L}_{j=1} \sigma^2_j \mathbold R_{L,j} + \sigma^2_{0} \mathbold I_n$, 
with the $(i,i')$th entry of the correlation matrix $\mathbold R_{L,j}$ being $\mathbold R_{L,j}(i,i')=c_{L,j}(\tilde x_{L,j,i}, \tilde x_{L,j,i'})$. 
For any input $\mathbold x^*$, the predictive distribution   conditional on the parameters and the transformed inputs of the final layer follows  
\begin{align}
(y(\mathbold x^*) \mid \mathbold y, \tilde{\mathbold x}_{L},\bmbold \beta_{L,1:d_L}, \bmbold \sigma^2_{0:d_L} ) \sim
\mathcal{N}\left(\hat y(\mathbold x^*), \,  \hat \sigma^2_y(\mathbold x^*) \right), 
\end{align} 
where the predictive mean and predictive variance follow
\begin{align}
\hat y(\mathbold x^*)&=  (\bmbold \Sigma^*_L)^T \bmbold{\tilde \Sigma}^{-1}_L
\mathbold y, \label{equ:pred_mean}\\
 \hat \sigma^2_y(\mathbold x^*) &= \sum^{d_L}_{j=1}\sigma^2_j+\sigma^2_{0}- (\bmbold \Sigma^*_L)^T \bmbold{\tilde \Sigma}^{-1}_L \bmbold \Sigma^*_L,
 \label{equ:sigma_2_y}
\end{align}
where  $\bmbold \Sigma^*_L=\sum^{d_L}_{j=1}\sigma^2_{j} \mathbold r^*_{L,j}$ and $\mathbold r^*_{L,j}=(c_{L,j}(\tilde x^*_{L,j}, \tilde x_{L,j,1}),...,c_{L,j}(\tilde x^*_{L,j}, \tilde x_{L,j,n}))^T$.
 
\subsection{Jointly inverse Kalman filter for fast computation}
\label{subsec:JIKF}
 We suppress the layer index  $l$ in this subsection. 
 Consider  a continuous-time DLM for an observation $ \tilde y_i=y(\tilde x^{(s)}_i)$ evaluated at a  non-decreasing scalar input sequence $\tilde x^{(s)}_i$  \cite{petris2009dynamic}
\begin{align}
 \tilde y_i&=\mathbold F  {\bmbold \theta}_i+ \tilde \epsilon_i, \label{equ:dlm_obs} \\
{\bmbold \theta}_i&= \mathbold G_i {\bmbold \theta}_{i-1}+ \mathbold {\tilde w}_i, 
\label{equ:dlm_states}
\end{align}
where $\mathbold F$  is a $q$-dimensional row vector that maps  the latent state vector ${\bmbold \theta}_i=\bmbold \theta(\tilde x^{(s)}_i)$  to  $ \tilde y_i$, $\tilde \epsilon_i \sim \mathcal{N}(0,\tilde \sigma^2_0)$,  $\mathbold G_i$ is a $q\times q$ transition matrix, 
and  $\mathbold{\tilde w}_i\sim \mathcal{N}(\mathbold 0, \mathbold{\tilde W}_i)$  for $i=2,...,n$. Assume the initial state vector follows the stationary distribution  ${\bmbold  \theta}_i \sim \mathcal{N}(\mathbold 0, \mathbold{\tilde W}_1)$. 
For the half-integer Mat{\'e}rn kernels considered here, the terms $\mathbold F$, $\mathbold G_i$ and $\tilde{\mathbold W}_i$ have closed-form expressions. 
For instance, $F=1$ for exponential kernel in Equation (\ref{equ:exp}) and $\mathbold F=(1,0,0)$ for Mat{\'e}rn kernel  in Equation (\ref{equ:matern_5_2}). The remaining expressions of the parameters are provided in Section S1 
of the {Supplementary Material}.

Our strategy is to scalably compute  $\mathbold b={ \bmbold \Sigma}  \mathbold v$ and its derivative with respect to parameters in GNets, where $\mathbold v=(v_1,...,v_n)^T$ is any $n$-dimensional  vector, and $\bmbold \Sigma=\mathrm{cov}[{\mathbold z}_{1:n}, {\mathbold z}_{1:n}]$ where $z_i=\mathbold F \bmbold \theta_i$ with latent state vector $\bmbold \theta_i$ follows Equation (\ref{equ:dlm_states}).

\begin{lemma}
The  covariance matrix between the latent states $\bmbold \theta_{i}$ and $\bmbold \theta_{i'}$  in Equation (\ref{equ:dlm_states}) follows 
\begin{align}
\bmbold \Sigma^{\theta}_{i,i'}&=\prod^{i}_{\tilde i=i'+1} \mathbold G_{\tilde i} \mathbold{\tilde W}_1, \, \mathrm{ when } \, i>i', \label{equ:Sigma_aug}  
\end{align}
with $\prod^{i}_{\tilde i=i'+1} \mathbold G_{\tilde i}=\mathbold G_{i}\mathbold G_{i-1}\cdots \mathbold G_{i'+1}$, for $i=2,...,n$, and $\bmbold \Sigma^{\theta}_{i,i} =\mathbold{\tilde W}_1$, for $i=1,...,n$.  Define 
\begin{align}
 \mathbold a^{-}_i&=\sum^{i}_{i'=1} v_{i'} \bmbold \Sigma^{\theta}_{i,i'} \mathbold F^T \label{equ:a_neg_direct}  \,\mathrm{ for }\, i=1,...,n, \\
 \mathbold a^{+}_{i}&=\sum^{n}_{i'=i+1} v_{i'} \bmbold \Sigma^{\theta}_{i',i} \mathbold F^T \label{equ:a_pos_direct}   \, \mathrm{ for } \, i=1,...,n-1,  \end{align}
and $\mathbold a^{+}_{n}=\mathbold 0_q$. Then, for $i=1,...,n$,  the $i$th entry of $\mathbold b=(b_{1},...,b_{n})^T={\bmbold \Sigma} \mathbold v$ follows
\begin{equation}
 b_{i}=  \mathbold F \mathbold a^{-}_i+ \mathbold F \mathbold a^{+}_i.
\label{equ:b_f}  
\end{equation}
\label{lemma:z_direct}
\end{lemma}

 To  compute $\mathbold b={ \bmbold \Sigma}  \mathbold v$,  the two quantities $\mathbold a_i^{-}$ and $\mathbold a_i^{+}$ in Lemma \ref{lemma:z_direct} collect the covariance contributions from indices up to $i$ and after $i$, respectively. This  decomposition is the key structure exploited for fast and exact computation in Theorem \ref{theorem:a_fast}. Furthermore, for time-homogeneous continuous-time process in (\ref{equ:dlm_states}), we notice that the transition matrices have the form $\mathbold G_i=e^{\mathbold J h_i}$ and are commutative:
\begin{equation}
\mathbold G_i \mathbold G_{i'}=  \mathbold G_{i'} \mathbold G_i.
\label{equ:G_commutative}
\end{equation}
As an example,  a GP with a half-integer Mat{\'e}rn kernel follows the state-space representation in Equations (\ref{equ:dlm_obs})-(\ref{equ:dlm_states}), and its transition matrices can be written as $\mathbold G_i=e^{\mathbold J (\tilde x^{(s)}_i-\tilde x^{(s)}_{i-1})}$ with $\mathbold J$ being a $q\times q$ matrix \cite{hartikainen2010kalman}, so  Equation (\ref{equ:G_commutative}) holds.

\begin{theorem}
 Assume that the transition matrix $\mathbold G_i$ satisfies the commutative property  in Equation (\ref{equ:G_commutative}).  Let $\mathbold a^{-}_1= v_1\bmbold \Sigma^{\theta}_{1,1} \mathbold F^T$ and  $\mathbold a^{+}_{n}=\mathbf 0_{q}$.
 Equations (\ref{equ:a_neg_direct}) and (\ref{equ:a_pos_direct}) can be iteratively computed by  
\begin{align}
\mathbold a^{-}_i&= \mathbold G_i \mathbold a^{-}_{i-1} +  v_i \mathbold{\tilde W}_1 \mathbold F^T,\, \mbox{ for } i= 2,...,n, 
\label{equ:a_neg_fast}\\
\mathbold a^{+}_i&= \mathbold G_{i+1} \mathbold a^{+}_{i+1} +v_{i+1}\mathbold G_{i+1} \mathbold{\tilde W}_1 \mathbold F^T,\, \mbox{ for } i=n-1,...,1.  
\label{equ:a_pos_fast}
\end{align}
\label{theorem:a_fast}
\end{theorem}

Together, Lemma \ref{lemma:z_direct} and Theorem \ref{theorem:a_fast} give an exact DLM-specific quasiseparable matrix--vector multiplication, reducing direct $\mathcal O(n^2)$ multiplication to $\mathcal O(nq^2)$ after sorting. Sorting requires a one-time $\mathcal O(n\log n)$ step and the order can be reused for subsequent products. Here $q=1$ for the exponential kernel in (\ref{equ:exp}) and $q=3$ for the Mat{\'e}rn kernel in (\ref{equ:matern_5_2}). In GNet this recursion is applied separately to every neuron for covariance matrix--vector products and their derivatives required by the model gradients. As $\tilde{\mathbold x}_{L,j}$ is a learned projection, different $\mathbold R_{L,j}$ generally require different sorting permutations. Thus $\tilde{\bmbold\Sigma}_L\mathbold v=\sigma_0^2\mathbold v+\sum_j\sigma_j^2\mathbold R_{L,j}\mathbold v$ is assembled from exact neuron-specific products rather than from one common one-dimensional ordering as in previous studies \cite{hartikainen2010kalman,nickisch2018state}.

\begin{figure}[t]
    \centering
        \begin{tabular}{ccc}
        \begin{overpic}[scale=0.37]{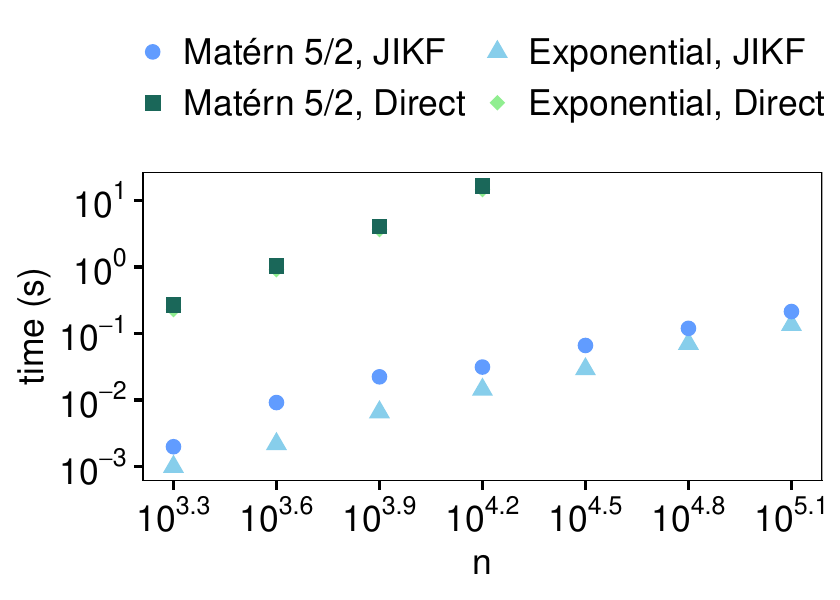}
            \put(0.5,65){{(a)}}
        \end{overpic} &
        \begin{overpic}[scale=0.37]{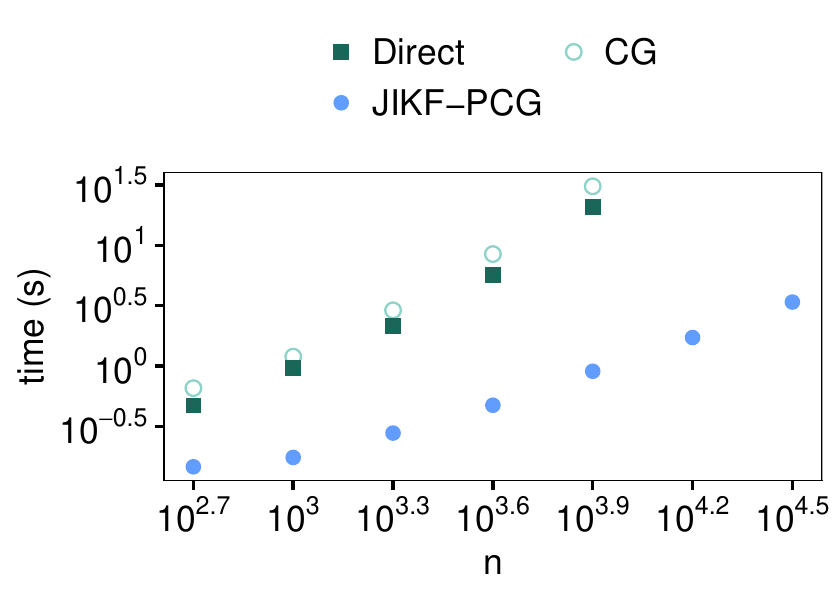}
            \put(0.5,65){{(b)}}
        \end{overpic}&
         \begin{overpic}[scale=0.34]{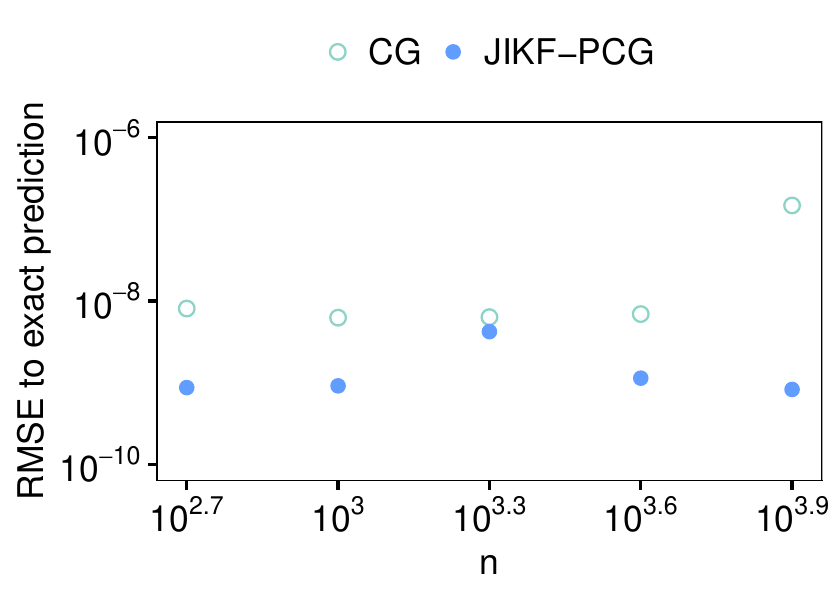}
            \put(0.5,71)
            {{(c)}} 
        \end{overpic}
        \end{tabular}
        \vspace{-.1in}
    \caption{(a) Computational time for $\mathbold b=\tilde{\bmbold \Sigma}_L \mathbold y$ by direct matrix-vector multiplication and by the JIKF for the  exponential kernel and Mat{\'e}rn kernel in Equations (\ref{equ:exp}) and (\ref{equ:matern_5_2}), respectively. Both methods are exact. The solid squares and diamonds overlap. 
    (b)  Time to compute $(\bmbold \Sigma^*_{L})^T \tilde{\bmbold \Sigma}^{-1}_L \mathbold y$ for predicting $n^*=10^4$ inputs  by direct computation, 
    conjugate gradient (CG) approach with direct matrix-vector multiplication, and JIKF-PCG.  In all panels, $d_L=10$ and  $p=2$. (c) The average root mean squared error (RMSE) between JIKF-PCG and direct computation is  smaller than $10^{-8}$ for the sample sizes where  direct computation is available. Each experiment is  repeated five times. 
     }
    \label{fig:computation_comparison}
\end{figure}

We call the iterative algorithm in Theorem \ref{theorem:a_fast} for  computing the state-space matrix-vector multiplication, the \textit{jointly inverse Kalman filter} (JIKF), which is oriented to accelerate prediction and model training for GNets. The name reflects its relation to the Kalman filter, which scalably computes $\mathbold L^{-1}\mathbold v$ where $\mathbold L$ is the Cholesky factor of $\bm\Sigma$. The state-space representation in  (\ref{equ:dlm_states})  leads to Equation (\ref{equ:Sigma_aug}), a crucial identity that enables fast covariance matrix--vector products. Compared to previous studies \cite{chandrasekaran2005some,nickisch2018state,fang2025inverse}, GNet introduces a different computational setting, where each neuron acts on a learned one-dimensional projection of potentially high-dimensional inputs and generally induces a different ordering. Compared with the inverse Kalman filter in \cite{fang2025inverse}, for instance, JIKF is much faster as it does not require running a separate Kalman filter first, which costs $\mathcal O(nq^3)$ in general. More importantly, JIKF accelerate both the covariance matrix-vector products and their derivatives with respect to model parameters, which are crucial for learning GP activations.

Figure \ref{fig:computation_comparison} (a) compares  JIKF with  direct matrix-vector multiplication, where the simulation details are provided in Section S3.4 of the Supplementary Material.
JIKF is substantially faster, and it gives exact results without approximation.  

To compute the predictive mean in Equation (\ref{equ:pred_mean}), we develop JIKF-PCG. Projected additive GPs have used approximate matrix--vector products inside CG \cite{delbridge2020randomly}, but these models do not learn multilayer neuron projections or use exact state-space covariance matrix-vector products to accelerate the computation. Here, each product is exact for the state-space kernels considered, and its projection is estimated. We use a Nystr{\"o}m preconditioner \cite{rudi2017falkon} to solve $\bmbold\alpha=\tilde{\bmbold\Sigma}_L^{-1}\mathbold y$, with JIKF applying $\tilde{\bmbold\Sigma}_L\mathbold v$ as a sum of neuron-specific products at every PCG iteration. We also accelerate $(\bmbold\Sigma_L^*)^T\bmbold\alpha$ by JIKF. After one-time processing, the cost of predicting each test input by a pretrained GNet grows only logarithmically with $n$, enabling fast repeated predictions. 
Computational details are given in Sections S3.1--S3.3
of the Supplementary Material.

 Figure \ref{fig:computation_comparison}   (b) shows that JIKF-PCG is substantially more scalable for computing the predictive mean than either direct computation or CG with direct matrix-vector products.  Figure \ref{fig:computation_comparison}(c) shows average RMSE below $10^{-8}$ relative to exact predictions, while Figure S1 
 in the Supplementary Material shows that Nystr{\"o}m preconditioning markedly reduces the number of iterations.

Exact evaluation of the predictive variance in Equation (\ref{equ:sigma_2_y}) at many test inputs remains costly. We therefore approximate its signal component using a small training subset, and calibrate predictive intervals with another held-out subset from training data. Section S3.5 of the Supplementary Material gives the procedure, and Section \ref{sec:numerical} evaluates its empirical coverage.

\section{Accelerated model training}
\label{sec:model_training}

We train GNets with Adam \cite{kingma2014adam}, although other optimization methods can be used.  
At each iteration, 
 the training data are partitioned into a  model building batch $\{\mathbold X^B, \mathbold y^B\}$ and a validation batch $\{\mathbold X^V, \mathbold y^V\}$, with the number of observations being $n_B$ and $n_V$, respectively.
The superscripts $B$ and $V$ mean the model building batch and the validation batch, respectively. 
We estimate the parameters by minimizing the predictive mean squared error
\begin{align}
\mathcal L=\frac{1}{n_{V}}\sum^{n_{V}}_{i=1}\left(y^V_i-\hat y^V_i  \right)^2, 
 \label{equ:Loss}
 \end{align} 
 where $ y^V_i=y(\mathbold x^{V}_{i})$ and $\hat y^V_i=\hat y(\mathbold x^{V}_{i})$ are the $i$th validation observation and prediction, respectively.

 Scalable GP calculations have been developed for structured kernels with 1D inputs \cite{foreman2017fast,nickisch2018state}, but they do not address gradients through multiple learned GP neurons and their trainable parameters. 
Lemmas \ref{lemma:gradients_one_layer} and \ref{lemma:general_nonfinal_layer_gradient} give closed-form gradients for the final and preceding layers. They avoid numerical differentiation and allow state-space covariance and derivative matrix-vector products to be accelerated by JIKF. In contrast, differentiating through repeated JIKF and JIKF-PCG iterations by automatic differentiation \cite{baydin2018automatic} would require large computational graphs and substantially higher storage costs.

We denote $\tilde \sigma^2_{j}=\log(\sigma^2_j)$  for  $j=0,...,d_L$, and $ \tilde \beta_{l,j'}=\log(\beta_{l,j'})$, for $j'=1,...,d_l$, and $l=1,...,L$.  Let $\bmbold{\tilde \Sigma}^B_L=\bmbold \Sigma^B_L+\sigma^2_0 \mathbold I_{n_B}$ where the  $n_B\times n_B$  matrix $\bmbold \Sigma^B_L= \sum^{d_L}_{j=1}\sigma^2_{j}\mathbold R^B_{L,j}$ with $\mathbold R^B_{L,j}$, 
and $\bmbold \Sigma^{V,B}_L=\sum^{d_L}_{j=1}\sigma^2_{j} \mathbold R^{V,B}_{L,j}  $ where the $(i,i')$th entry of $n_V\times n_B$ matrix $\mathbold R^{V,B}_{L,j} $  is $c_{L,j}(\tilde x^V_{L,j,i},\tilde x^B_{L,j,i'})$. 
Furthermore, for $C \in \{B, V\}$, let 
$\mathbold Z^C_{l-1}$  be an $n_C\times d_{l-1}$ matrix containing  activation function values at $l-1$th layer, and $\tilde{\mathbold x}^C_{l,j}={\mathbold Z}^C_{l-1} \mathbold w_{l,j}$ is a vector of $n_C$ dimensions.  Lemma \ref{lemma:gradients_one_layer} and  Lemma \ref{lemma:general_nonfinal_layer_gradient} apply to  differentiable kernels, including the   Mat{\'e}rn kernel in Equation (\ref{equ:matern_5_2}), which is used in Section \ref{sec:numerical}.  Extending the gradient formulas to kernels that are not differentiable at coincident inputs, such as that in Equation (\ref{equ:exp}), is left for future work. The proofs of these lemmas are available in Section S4
in the Supplementary Material.

\begin{lemma}[Gradients of a GNet at the final layer $L$]
Consider a GNet with differentiable kernels and  the loss function in  (\ref{equ:Loss}), let 
\begin{align}
  \mathbold u=(\bmbold{\tilde \Sigma}^B_L)^{-1}\mathbold y^B, \quad 
  \tilde{\mathbold u}=(\bmbold{\tilde \Sigma}^B_L)^{-1}(\bmbold \Sigma^{V,B}_L)^T{\bmbold \Delta_{y^V}}, \quad \mbox{ and } \quad 
  \bmbold \Delta_{y^V}=\frac{\hat{\mathbold y}^{V}-\mathbold y^V}{n_{V}}. \label{equ:u_v_Delta_y_star}
  \end{align}
\begin{enumerate}
\item The gradient for $\tilde {\sigma}^2_{0}$ and $\tilde {\sigma}^2_{j}$ for $j=1,...,d_L$ follows 
\begin{align}
\frac{\partial \mathcal L}{\partial  {\tilde \sigma}^2_{0}}&=-2 \exp(\tilde {\sigma}^2_{0}) \tilde{\mathbold u}^T \mathbold u, \label{equ:grad_sigma_2_0_L}\\
\frac{\partial \mathcal L}{\partial  {\tilde \sigma}^2_{j}}&= 2 \exp(\tilde {\sigma}^2_{j}) (\bmbold \Delta_{y^V}^T \mathbold R^{V,B}_{L,j} \mathbold u - \tilde{\mathbold u}^T\mathbold R^B_{L,j} \mathbold u  ).
\label{equ:grad_sigma_2_j_L}
\end{align}
\item The gradient of $\tilde \beta_{L,j}$, for $j=1,...,d_L$, follows
\begin{align}
\frac{\partial \mathcal L}{\partial \tilde \beta_{L,j}} 
=&  2\exp(\tilde\sigma_j^2)\exp(\tilde \beta_{L,j}) \left(\bmbold \Delta_{y^V}^T  D_{\beta_{L,j}}\mathbold R^{V,B}_{L,j} \mathbold u  - \tilde{\mathbold u}^T D_{\beta_{L,j}}\mathbold R^B_{L,j} \mathbold u  \right), 
\label{equ:grad_beta_j_L}
\end{align} 
where $D_{\beta_{L,j}}\mathbold R^{V,B}_{L,j} \mathbold u=({\partial \mathbold R^{V,B}_{L,j}}/{\partial   \beta_{L,j}}) \mathbold u$ and $D_{\beta_{L,j}}\mathbold R^B_{L,j} \mathbold u=({\partial \mathbold R^B_{L,j}}/{\partial   \beta_{L,j}}) \mathbold u$.
\item The gradient of $\mathbold w_{L,j}$  
for $j=1,...,d_L$   follows  
\begin{equation}
  \frac{\partial \mathcal L}{\partial \mathbold w_{L,j}}
  =
 ({\mathbold Z}^{B}_{L-1})^T   \tilde{\mathbold g}^{B}_{L,j} 
  +
 ({\mathbold Z}^{V}_{L-1})^T   \tilde{\mathbold g}^{V}_{L,j}, 
  \qquad j=1,\ldots,d_L,
  \label{equ:grad_w_matrix_L_1}
\end{equation}
where $ \tilde{\mathbold g}^V_{L,j} = {\partial \mathcal L}/{\partial \tilde{\mathbold x}^{V}_{L,j} }$ and $ \tilde{\mathbold g}^B_{L,j} ={\partial \mathcal L}/{\partial  \tilde{\mathbold x}^{B}_{L,j} }$ are given by
\begin{align}
  \tilde{\mathbold g}^{B}_{L,j}&= 2 \exp(\tilde\sigma_j^2)
  \left\{\mathbold u \circ
  \left[D_2(\mathbold R^{V,B}_{L,j})^T \bmbold \Delta_{y^V} \right] -
\tilde{\mathbold u} \circ (D_1\mathbold R^B_{L,j}\mathbold u)-
  \mathbold u \circ (D_1\mathbold R^B_{L,j}\tilde{\mathbold u})
  \right\}, 
    \label{equ:g_train} \\
    \tilde{\mathbold g}^{V}_{L,j}
 & =
  2\exp(\tilde\sigma_j^2)\,
  \bmbold \Delta_{y^V} \circ
  \left(D_1\mathbold R^{V,B}_{L,j}\mathbold u\right),
  \label{equ:g_val}
  \end{align} 
  with $\circ$ denoting the elementwise product. Here for any  $n_1\times n_2$ matrix $\mathbold M$ with $M_{i,i'}=c(x_{1,i}, x_{2,i'})$, any two vectors $\mathbold v$ and  $\tilde{\mathbold v}$ of sizes $n_1$ and $n_2$, respectively, 
each entry of  the row-wise and column-wise derivatives of the matrix-vector product is 
\begin{align*}
  (D_1\mathbold M\tilde{\mathbold v})_i=
  \sum^{n_2}_{i'=1}
  \frac{\partial c(x_{1,i},x_{2,i'})}{\partial x_{1,i}}\tilde v_{i'}, \quad  \mbox{ and }  \quad (D_2 \mathbold M^T\mathbold v)_{i'}=
  \sum^{n_1}_{i=1}
  \frac{\partial  c(x_{1,i},x_{2,i'})}{\partial x_{2,i'}}v_{i}, 
\end{align*}
for $i=1,...,n_1$ and $i'=1,...,n_2$. 
\end{enumerate}
\label{lemma:gradients_one_layer}
\end{lemma}

For a one-layer GNet, i.e. $L=1$, Lemma \ref{lemma:gradients_one_layer} gives all parameter gradients, with $ {\mathbold Z}^{B}_{0} =\mathbold X^B $  and $ {\mathbold Z}^{V}_{0} =\mathbold X^V $. The covariance and derivatives of matrix-vector products in Equations (\ref{equ:grad_beta_j_L}), and (\ref{equ:g_train})-(\ref{equ:g_val}) can be accelerated by JIKF. 
For $L>1$, directly marginalizing out the latent GPs  $z_{l,j}(\cdot)$ is generally intractable. 
For each  neuron at an intermediate layer, let $\mathbold z^{K}_{l,j}$  be its $m_{l,K}$ knot values, with prespecified 1D knots inputs $\tilde {\mathbold x}^{K}_{l,j}$, 
where the superscript $K$ denotes knots, and denote  $\mathbold R^K_{l,j}$ the corresponding knot correlation matrix, with the $(i,i')$th entry  $c_{l,j}(\tilde x^K_{l,j,i}, \tilde x^K_{l,j,i'})$. 
For $C\in\{B,V\}$, define $\mathbold R^{C,K}_{l,j}$ with $(i,i')$th entry $c_{l,j}(\tilde x^C_{l,j,i},\tilde x^K_{l,j,i'})$. Then, for $l=1,...,L-1$, 
\begin{equation}
\mathbold z^{C}_{l,j}= \mathbold R^{C,K}_{l,j} \bmbold \alpha^K_{l,j},
\label{equ:z_C_l_j}
\end{equation}
 where $\bmbold \alpha^K_{l,j}=(\mathbold L^K_{l,j})^{-T} \mathbold z^{K,W}_{l,j}$ and   $\mathbold z^{K,W}_{l,j}=  ({\mathbold L}^K_{l,j})^{-1} \mathbold z^{K}_{l,j}$,   with $ {\mathbold L}^K_{l,j}$ being the  Cholesky factor of $\mathbold R^K_{l,j}= {\mathbold L}^K_{l,j} ({\mathbold L}^K_{l,j})^T$, which does not need to be computed directly. We optimize the whitened knot values   $\mathbold z^{K,W}_{l,j}$.  As  input in each neuron  $\tilde x_{l,j,i}$ is 1D, a small knot set works well. In  all numerical studies, $m_{l,K}=10$ is used for $1\leq l\leq L-1$, and $m_{L,K}=0$ as no knot is required for the final layer.

\begin{lemma}[Gradients of a GNet at any layer $1\leq l<L$ with $L\geq 2$]
\label{lemma:general_nonfinal_layer_gradient}
Consider a GNet with differentiable kernels and the loss function  in Equation (\ref{equ:Loss}). 
For any non-final layer $l<L$ and any
component $j=1,\ldots,d_l$, we have 
\begin{align}
  \frac{\partial \mathcal L}{\partial \mathbold w_{l,j}}
  &= (\mathbold Z_{l-1}^{B})^T\tilde{\mathbold g}_{l,j}^{B}
     +(\mathbold Z_{l-1}^{V})^T\tilde{\mathbold g}_{l,j}^{V},
  \label{equ:general_grad_w_lj}\\
  \frac{\partial \mathcal L}{\partial\tilde\beta_{l,j}}
  &= \exp(\tilde\beta_{l,j})\sum_{C\in \{B,V\}}\left[
     (\mathbold g_{l,j}^{C})^T
     \left(D_{\beta_{l,j}}\mathbold R_{l,j}^{C,K}\bmbold\alpha^K_{l,j}+\mathbold R_{l,j}^{C,K}D_{\beta_{l,j}}{\bmbold \alpha}^K_{l,j}\right)
     \right],
  \label{equ:general_grad_beta_lj}\\
  \frac{\partial \mathcal L}{\partial{\mathbold z}_{l,j}^{K,W}}
  &=(\mathbold L_{l,j}^{K})^{-1} 
     \left[(\mathbold R_{l,j}^{B,K})^T\mathbold g_{l,j}^{B}
     +(\mathbold R_{l,j}^{V,K})^T\mathbold g_{l,j}^{V}\right],
  \label{equ:general_grad_eta_lj}
\end{align}
where  $D_{\beta_{l,j}}{\bmbold\alpha}^K_{l,j}=\partial \bmbold \alpha^{K}_{l,j}/\partial  \beta_{l,j}$, and for any $C \in \{B, V\}$,   $\tilde{\mathbold g}_{l,j}^C={\partial \mathcal L}/{\partial \tilde{\mathbold x}_{l,j}^C}$ and $\mathbold g_{l,j}^C={\partial \mathcal L}/{\partial\mathbold z_{l,j}^C}$ follow
\begin{align}
\tilde{\mathbold g}_{l,j}^C
  &  =\mathbold g_{l,j}^C\circ
  \left(D_1\mathbold R_{l,j}^{C,K}\bmbold\alpha^K_{l,j}\right),  \label{equ:tilde_g_lj}
 \\
  \mathbold g_{l,j}^C
  &=\sum_{j'=1}^{d_{l+1}} w_{l+1,j',j} \tilde{\mathbold g}_{l+1,j'}^C, \label{equ:general_backprop_adjoint}
\end{align}
for $l=1,...,L-1$, and   $\tilde{\mathbold g}^B_{L,j}$ and $\tilde{\mathbold g}^V_{L,j}$ in the final layer are defined in (\ref{equ:g_train}) and (\ref{equ:g_val}), respectively. 
\end{lemma}

All  products between state-space covariances and vectors are accelerated by  JIKF or JIKF-PCG during model training. 
These include for instance, computing   $\mathbf u$ and $\tilde{\mathbf u}$ in    (\ref{equ:u_v_Delta_y_star}) and the products of the form 
$D_{\beta_{L,j}}\mathbold R^{V,B}_{L,j} \mathbold u$, $D_{\beta_{L,j}}\mathbold R^B_{L,j} \mathbold u$  in  (\ref{equ:grad_beta_j_L}), without explicitly forming  covariance matrices. 
The computational details 
are summarized in Section 
S5 
of the Supplementary Material, and Table S1 summarizes the overall computational complexity. The JIKF matrix-vector products are linear after sorting, while including sorting gives near-linear dependence on the minibatch size. After pretraining, each prediction increases only logarithmically with the training sample size.

\section{Theoretical studies}
\label{sec:theory}

\subsection{Function classes of GNets}
\label{sec:function_class}

We compare the function classes of GNet and the feedforward neural network (FNN) under a fixed architecture. Let $\mathcal X\subset\mathbb R^p$ be compact and let $C(\mathcal X)$ denote the real-valued continuous functions on $\mathcal X$, with $\|f\|_{\infty,\mathcal X}=\sup_{\mathbold x\in\mathcal X}|f(\mathbold x)|$. Although FNNs with a fixed nonpolynomial activation and growing width can be dense in $C(\mathcal X)$ \cite{leshno1993multilayer}, here we compare function classes with  the fixed number of layers $L$ and neurons $\mathbold d=(d_0,\ldots,d_L)$, where $d_0=p$.

We consider the finite kernel-expansion class induced by the GNet activation kernels below
\begin{align*}
\mathcal G(L, \mathbold d)=\Bigl\{f_G(\mathbold x)=\sum_{j=1}^{d_L}\sigma_jz_{L,j}(\mathbold x):\,
&z_{l,j}(\mathbold x)=g_{l,j}\left(\mathbold w_{l,j}^T\mathbold z_{l-1}(\mathbold x)\right),\,
 g_{l,j}(t)=\sum_{i=1}^{n_{l,j}}\alpha_{l,j,i}c_{l,j}(t,t_{l,j,i})
\Bigr\},
\end{align*}
where $n_{l,j} \in \mathbb N^+$, $t,t_{l,j,i},\alpha_{l,j,i}\in\mathbb R$, $\mathbold w_{l,j}\in\mathbb R^{d_{l-1}}$, and $\sigma_j\in\mathbb R^+$. At the initial layer, $\mathbold z_0(\mathbold x)=\mathbold x$ and $d_0=p$. Let $\overline{\mathcal G}(L, \mathbold d)$ denote its closure in $C(\mathcal X)$ under $\|\cdot\|_{\infty,\mathcal X}$.

For a fixed activation function $\psi$, define the corresponding FNN function class by
\begin{align*}
\mathcal F_{\psi}(L, \mathbold d)=\Bigl\{f_F:\;
 f_F(\mathbold x)=\sum_{j=1}^{d_L}a_jz^F_{L,j}(\mathbold x), \quad
 z^F_{l,j}(\mathbold x)=\psi\left\{(\mathbold w^F_{l,j})^T\mathbold z^F_{l-1}(\mathbold x)+b_{l,j}\right\}
\Bigr\},
\end{align*}
where $\mathbold w^F_{l,j}\in\mathbb R^{d_{l-1}}$, $b_{l,j},a_j\in\mathbb R$, and $\mathbold z^F_0(\mathbold x)=\mathbold x$. Let $\overline{\mathcal F}_{\psi}(L, \mathbold d)$ denote its closure under $\|\cdot\|_{\infty,\mathcal X}$.

\begin{assumption}
\label{assumption:function_class}
\begin{enumerate}
\renewcommand{\theenumi}{A\arabic{enumi}}\renewcommand{\labelenumi}{(\theenumi)}
\item \label{assump:fc_compact} (Input space and fixed architecture). The input space $\mathcal X\subset\mathbb R^p$ is compact. The number of  layers and neurons $(L,\mathbold d)$ are finite and fixed.
\item \label{assump:fc_activation} (Activation function). The activation $\psi:\mathbb R\to\mathbb R$ is continuous and locally Lipschitz.
\item \label{assump:fc_kernel} (Kernel). Each kernel $c_{l,j}$ is continuous and strictly positive definite on $\mathbb R$, and universal on compact intervals for every compact interval $\mathcal I\subset\mathbb R$, meaning that for every continuous function $f:\mathcal I\to\mathbb R$, and every $\epsilon_c>0$, there exist a finite integer $n_0$, centers $t_1,\ldots,t_{n_0}\in\mathcal I$, and coefficients $\alpha_1,\ldots,\alpha_{n_0}\in\mathbb R$ such that
\begin{align}
\sup_{t\in\mathcal I}\left|f(t)-\sum_{i=1}^{n_0}\alpha_i c_{l,j}(t,t_i)\right|<\epsilon_c.
\label{equ:kernel_universal_assumption}
\end{align}
\end{enumerate}
\end{assumption}
Commonly used activations, including ReLU, softplus, and  GELU \cite{hendrycks2016gelu}, satisfy (\ref{assump:fc_activation}). Translation invariant kernels with positive spectral densities satisfy (\ref{equ:kernel_universal_assumption}) \cite{micchelli2006universal}, and this includes the half-integer Mat{\'e}rn kernels in (\ref{equ:exp}) and (\ref{equ:matern_5_2}).

\begin{lemma}[Approximation of an FNN by a GNet]
\label{lemma:gnet_contains_fnn}
Under Assumption \ref{assumption:function_class}, for every $f_F\in\mathcal F_{\psi}(L, \mathbold d)$ and $\epsilon_f>0$, there exists $f_G\in\mathcal G(L, \mathbold d)$ with the same $L$ and $\mathbold d$ such that $\|f_F-f_G\|_{\infty,\mathcal X}<\epsilon_f$.
\end{lemma}

\begin{corollary}[Containment of the FNN class with a fixed architecture]
\label{corollary:no_fnn_outside_gnet}
Under Assumption \ref{assumption:function_class}, for any $\psi$ satisfying (\ref{assump:fc_activation}), including ReLU, softplus, and GELU, 
\begin{align}
\overline{\mathcal F}_{\psi}(L, \mathbold d)
\subseteq
\overline{\mathcal G}(L, \mathbold d).
\label{equ:no_fnn_outside_gnet}
\end{align}
\end{corollary}

The comparison in Lemma \ref{lemma:gnet_contains_fnn} and Corollary \ref{corollary:no_fnn_outside_gnet} fix the numbers of layers and neurons but not the number of kernel terms in GNet. The following example shows that the containment can be strict for a fixed one-layer architecture.

\begin{example}[Strict separation for one-layer GNet and one-hidden-layer FNN with ReLU activation]
\label{example:gnet_strict_relu}
Let $\mathcal X=[0,1]^p$ and consider the inputs of the form  $\mathbold x=(x_1,0,\ldots,0)^T$ with $x_1\in[0,1]$. Set the number of separation nodes $m_{\rm sep}=d_L+3$, $t_i=i/(m_{\rm sep}+1)$, and $y_i=(-1)^i$, for $i=1,\ldots,m_{\rm sep}$. Let $c=c_{1,1}$ be the first-neuron kernel, and define $\mathbold R_{\rm sep}=\{c(t_i,t_{i'})\}_{i,i'=1}^{m_{\rm sep}}$,
$\mathbold r_{\rm sep}(x_1)=(c(x_1,t_1),\ldots,c(x_1,t_{m_{\rm sep}}))^T$, and 
$\mathbold y_{\rm sep}=(y_1,\ldots,y_{m_{\rm sep}})^T$. 
By strict positive definiteness, $\mathbold R_{\rm sep}$ is invertible. Let the target function be
\begin{align}
f_{\rm sep}(\mathbold x)=\mathbold r_{\rm sep}(x_1)^T\mathbold R_{\rm sep}^{-1}\mathbold y_{\rm sep},
\label{equ:gnet_separation_target}
\end{align}
which satisfies $f_{\rm sep}(t_i,0,\ldots,0)=(-1)^i$ and belongs to $\mathcal G(1, \mathbold d)$ with $\mathbold w_{1,1}=(1,0,\ldots,0)^T$, $n_{1,1}=m_{\rm sep}$, $t_{1,1,i}=t_i$, $\alpha_{1,1,i}$ being the $i$th entry of $\mathbold R_{\rm sep}^{-1}\mathbold  y_{\rm sep}$,  $\sigma_1=1$, and remaining neurons being zero.

In contrast, the restriction of any $f_F\in\mathcal F_{\rm ReLU}(1, \mathbold d)$ to this line is continuous piecewise affine with at most $d_L$ breakpoints. 

The derivation in Section S6 
in the Supplementary Material gives 
\begin{align}
\inf_{f_F\in\overline{\mathcal F}_{\rm ReLU}(1, \mathbold d)}
\|f_F-f_{\rm sep}\|_{\infty,\mathcal X}\ge 1.
\label{equ:relu_lower_bound}
\end{align}
\end{example}

\begin{corollary}[Strict one-layer inclusion with a fixed architecture]
\label{corollary:strict_one_layer_inclusion}
Under Assumption \ref{assumption:function_class} and $\mathcal X=[0,1]^p$, for every fixed $d_L<\infty$,
\begin{align}
\overline{\mathcal F}_{\rm ReLU}(1, \mathbold d)
\subsetneq
\overline{\mathcal G}(1, \mathbold d).
\label{equ:strict_function_class_inclusion}
\end{align}
In particular, this applies to the half-integer Mat{\'e}rn kernels used in GNet.
\end{corollary}

This strict inclusion in Corollary \ref{corollary:strict_one_layer_inclusion} does not conflict with universal approximation by FNNs of growing width \cite{leshno1993multilayer}, because the separating target in Example \ref{example:gnet_strict_relu} depends on the one layer and a fixed number of neurons $d_L$.

\subsection{Predictive risk bounds}
\label{sec:gnet_convergence_rates}

We study the statistical prediction risk of one-layer GNet in this section. Throughout this subsection, assume the fixed input dimension $p\ge2$ and consider the true data generating process
\begin{equation}
Y_i=f_0(\mathbold X_i)+\varepsilon_i,\qquad
\label{equ:theory_regression_model}
\end{equation}
where $\mathbold X_i\stackrel{\mathrm{i.i.d.}}{\sim}P_X$ and, independently,
$\varepsilon_i\stackrel{\mathrm{i.i.d.}}{\sim}\mathcal N(0,\sigma_{0,*}^2)$ with
$\sigma_{0,*}^2>0$ for $i=1,\ldots,n$. In this section, let $\mathcal X=[-1,1]^p$, and assume that $P_X$ has density $p_X$ satisfying
\begin{align}
0<p_{\min}\le p_X(\mathbold x)\le p_{\max}<\infty. 
\label{equ:gnet_random_design_density}
\end{align}

Split the data into independent model-building and validation samples
$\mathcal D_B=\{(\mathbold X_i^B,Y_i^B)\}_{i=1}^{n_B}$ and
$\mathcal D_V=\{(\mathbold X_i^V,Y_i^V)\}_{i=1}^{n_V}$, where
$n_B+n_V=n$ and $n_B$ and  $n_V$ grow as the same order as $n$, i.e., $n_B\asymp n_V\asymp n$. This data  split is used for  theoretical studies and it differs from the minibatches in Section~\ref{sec:model_training} and Section~\ref{sec:numerical}. 
 Define
\begin{align}
\|f\|_P^2=\int_{\mathcal X}f(\mathbold x)^2\,dP_X(\mathbold x).
\label{equ:norm_P}
\end{align}

Normalize $\|\mathbold w_j\|_2=1$ and let $\mathbb S^{p-1}=\{\mathbold w\in\mathbb R^p:\|\mathbold w\|_2=1\}$. 
 Write
$\vartheta_c=(\mathbold W,\bmbold\beta,\bmbold\tau)$, where
$\mathbold W=(\mathbold w_1,\ldots,\mathbold w_{d_L})\in(\mathbb S^{p-1})^{d_L}$,
$\bmbold\beta=(\beta_1,\ldots,\beta_{d_L})^T$, and
$\bmbold\tau=(\tau_1,\ldots,\tau_{d_L})^T$ with
$\tau_j=\sigma_j^2/\sigma_G^2$ and $\sigma_G^2=\sum_{j=1}^{d_L}\sigma_j^2$. Let
$\mathcal I_p=[-\sqrt p,\sqrt p]$, which contains $\mathbold w^T_j\mathbold x$ for
$\mathbold w_j\in\mathbb S^{p-1}$ and $\mathbold x\in\mathcal X$. Define
\begin{align}
\bar c_{\vartheta_{c}}(\mathbold x,\mathbold x')
=
\sum_{j=1}^{d_L}\tau_j
c_{\beta_j}(\mathbold w_j^T\mathbold x,\mathbold w_j^T\mathbold x').
\label{equ:gnet_pm_normalized_kernel}
\end{align}
where $c_{\beta_j}$ denotes the kernel with inverse range $\beta_j>0$ and one-dimensional inputs. 
Let $\bar{\mathbold R}_{\vartheta_{c}}^B$ and
$\bar{\mathbold r}_{\vartheta_{c}}^B(\mathbold x)$ denote the corresponding Gram matrix and cross-kernel vector on $\mathcal D_B$.

Assume $\|f_0\|_{\infty,\mathcal X}\le f_{0,max}$ and define 
$g_{f_{max}}(z)=\max\{-f_{max},\min(f_{max},z)\}$ for a fixed positive bound $f_{max}>f_{0,max}$. The bounded posterior mean estimator with a regularization parameter $\lambda$ is
\begin{align}
\hat f_{\vartheta}^{B}(\mathbold x)
=
g_{f_{max}}\left\{
\left(\bar{\mathbold r}_{\vartheta_{c}}^{B}(\mathbold x)\right)^{T}
\left(\bar{\mathbold R}_{\vartheta_{c}}^B+n_B\lambda\mathbold I_{n_B}\right)^{-1}
\mathbold Y^B
\right\}.
\label{equ:gnet_pm_krr_form}
\end{align}
This is the normalized GNet posterior mean in (\ref{equ:pred_mean}) with
$n_B\lambda=\sigma_0^2/\sigma_G^2$, where $\sigma_0^2$ is the model nugget variance;
$\sigma_{0,*}^2$ in (\ref{equ:theory_regression_model}) denotes the true noise variance.
Fix $0<\beta_{\min}<\beta_{\max}<\infty$ and $0<C_\lambda<1$. Define
$\vartheta=(\vartheta_c,\lambda)=(\mathbold W,\bmbold\beta,\bmbold\tau,\lambda)$  and its parameter space
\begin{align}
\Theta_{d_L}
=
\left\{\vartheta:
\mathbf W\in(\mathbb S^{p-1})^{d_L},
\bmbold\beta\in[\beta_{\min},\beta_{\max}]^{d_L},
\bmbold\tau\in\Delta_{d_L-1},
\lambda\in[C_\lambda/n_B,1]
\right\}.
\label{equ:gnet_pm_parameter_set}
\end{align}
where $\Delta_{d_L-1}=\{\bmbold\tau\in\mathbb R^{d_L}:\tau_j\ge0,\ \sum_{j=1}^{d_L}\tau_j=1\}$ is a probability simplex.
The constrained space has $d_{\Theta}=d_L(p+1)$ free continuous parameters: $d_L(p-1)$ direction coordinates, $d_L$ inverse ranges, $d_L-1$ simplex coordinates, and one regularization parameter. The corresponding unconstrained GNet parameterization has $d_L(p+2)+1$ parameters. The constraints are used only for the theory and need not be imposed in computation.

For a fixed width $d_L$, let $\hat\vartheta\in\Theta_{d_L}$ be the estimated parameters  and define
\begin{align}
\mathcal L(\vartheta)
&=
\frac1{n_V}\sum_{i=1}^{n_V}
\{Y_i^V-\hat f_{\vartheta}^{B}(\mathbold X_i^V)\}^2,
\label{equ:gnet_pm_validation_loss_theory}\\
\mathcal R_{n_B,d_L}^{\rm PM}(f_0)
&=
\inf_{\vartheta\in\Theta_{d_L}}
\mathbb E_{\mathcal D_B}\|\hat f_{\vartheta}^{B}-f_0\|_P^2,
\label{equ:gnet_pm_oracle_risk}\\
\mathcal R_{n,d_L}^{\rm opt}(f_0)
&=
\mathbb E_{\mathcal D,\rm alg}\left[
\mathcal L(\hat\vartheta)
-
\inf_{\vartheta\in\Theta_{d_L}}\mathcal L(\vartheta)
\right], 
\label{equ:gnet_pm_optimization_gap}
\end{align}
where the superscripts ``PM'' and ``opt'' denote posterior-mean oracle risk and optimization error. 
Here  the first expectation averages only over $\mathcal D_B$, and the second expectation averages over the full data and any randomness in the optimization algorithm. 

Prediction risk decompositions for neural-network estimators include, for instance, Lemma~4 of \cite{schmidt2020nonparametric}. The bounded predictive mean estimator of GNet  in (\ref{equ:gnet_pm_krr_form}), however, is substantially different from standard neural network predictors, and thus  new techniques are required for studying its convergence rates.
Theorem \ref{theorem:gnet_bound_decomposition} decomposes prediction risk into the posterior-mean oracle risk,  parameter  complexity, and optimization error.

\begin{theorem}[Posterior-mean prediction risk]
\label{theorem:gnet_bound_decomposition}
Assume that $\{c_{\beta}:\beta\in[\beta_{\min},\beta_{\max}]\}$ is a positive-semidefinite correlation-kernel family on one-dimensional inputs, uniformly bounded on $\mathcal I_p^2$ and uniformly Lipschitz jointly in its two inputs and inverse-range parameter. Under
(\ref{equ:theory_regression_model}) and $\|f_0\|_{\infty,\mathcal X}\le f_{0,max}$, for any fixed $a_d>0$ and sufficiently large $n$,
\begin{align}
\mathbb E_{\mathcal D,\rm alg}
\|\hat f_{\hat\vartheta}^{B}-f_0\|_P^2
\le
C\left\{
\mathcal R_{n_B,d_L}^{\rm PM}(f_0)
+
\frac{d_{\Theta}\log n}{n_V}
+
\mathcal R_{n,d_L}^{\rm opt}(f_0)
\right\}
+\mathcal O(n^{-2}),
\label{equ:gnet_pm_stat_opt_bound}
\end{align}
uniformly over $d_L\le n^{a_d}$. Here $C$ may depend on the fixed quantities in the setup, including $p$, $a_d$, $f_{0,max}$, $f_{max}$, $\sigma_{0,*}$, $C_\lambda$, $\beta_{\min}$, $\beta_{\max}$, and the kernel bounds, but not on $n$ or $d_L$.
\end{theorem}
The kernel conditions hold for all half-integer Mat{\'e}rn kernels with inverse ranges in $[\beta_{\min},\beta_{\max}]$, without requiring smoothness of $f_0$.  The three terms in (\ref{equ:gnet_pm_stat_opt_bound}) respectively quantify the statistical oracle risk, parameter-selection complexity, and optimization error.

For $s>0$, let $H^s(\mathcal I)$ denote the $L_2$-Sobolev space of order $s$ at domain $\mathcal I$ and define
\begin{align}
\mathcal W_{p}^{s}
=
\Big\{f\in H^s(\mathcal X):
\|f\|_{H^s(\mathcal X)}\le\delta_{\mathcal R},\ 
\|f\|_{\infty,\mathcal X}\le f_{0,max}
\Big\}.
\label{equ:bounded_sobolev_ball}
\end{align}
For noninteger $s$, $H^s(\mathcal X)$ denotes the restriction to $\mathcal X$ of $H^s(\mathbb R^p)$. For fixed width $d_L$, consider the ridge class
\begin{align}
\mathcal W_{\rm ridge}^{s}=
\Big\{f:\;&
f(\mathbold x)=\sum_{j=1}^{d_L}g_j(\mathbold w_j^T\mathbold x),\quad
\mathbold w_j\in\mathbb S^{p-1},\quad
g_j\in H^s(\mathcal I_p),\nonumber\\
&\sum_{j=1}^{d_L}\|g_j\|_{H^s(\mathcal I_p)}^2\le\delta_{\mathcal R}^2,\quad
\sum_{j=1}^{d_L}\|g_j\|_{\infty,\mathcal I_p}\le f_{0,max}
\Big\}.
\label{equ:smooth_ridge_class}
\end{align}
For the remaining rate results, suppose $c_\beta$ is the half-integer Mat{\'e}rn-$\nu$ correlation kernel, where $\nu=(2q-1)/2$ and $q\in\mathbb N^+$.

\begin{theorem}[The oracle rate of posterior mean for fixed-width ridge functions]
\label{theorem:gnet_ridge_minimax}
Under (\ref{equ:theory_regression_model})--(\ref{equ:gnet_random_design_density}), fix $p$ and $d_L$ and assume
\begin{align}
\nu\le s\le q=\nu+\frac12.
\label{equ:gnet_ridge_smoothness_window}
\end{align}
Then
\begin{align}
\sup_{f_0\in\mathcal W_{\rm ridge}^{s}}
\mathcal R_{n_B,d_L}^{\rm PM}(f_0)
\lesssim
n_B^{-\frac{2s}{2s+1}}.
\label{equ:gnet_ridge_pm_oracle_rate}
\end{align}
\end{theorem}
For fixed $p$ and $d_L$, the selection term in Theorem~\ref{theorem:gnet_bound_decomposition} is $O(\log n/n)$, and therefore smaller than $n^{-2s/(2s+1)}$. If the optimization error satisfies $\mathcal R_{n,d_L}^{\rm opt}(f_0)=O\{n^{-2s/(2s+1)}\}$, the selected GNet has the same rate as the standard one-dimensional Sobolev rate, reflecting the ridge structure in (\ref{equ:smooth_ridge_class}). 
Optimization-error rate for GNets remains an open problem.

Theorem \ref{theorem:gnet_pm_generic_rate} next allows the number of neurons to increase with sample size.
\begin{theorem}[Convergence rate for smooth functions]
\label{theorem:gnet_pm_generic_rate}
Under (\ref{equ:theory_regression_model})--(\ref{equ:gnet_random_design_density}), assume
$f_0\in\mathcal W_{p}^{s}$ and
\begin{align}
\nu\le s\le\nu+\frac p2.
\label{equ:gnet_pm_minimax_window}
\end{align}
For a sufficiently large fixed constant $C_D$, independent of $s$, denote a candidate set of  neuron counts with $\mathcal O(\log(n_B))$ elements
\begin{align}
\mathcal M_{n_B}
=
\left\{2^k:
0\le k\le
\left\lceil
\log_2\!\left(C_D n_B^{\frac{p-1}{2\nu+p}}\right)
\right\rceil
\right\}.
\label{equ:gnet_pm_candidate_widths}
\end{align}
For sufficiently large fixed $\kappa>0$, also independent of $s$, let $d\in\mathcal M_{n_B}$ and $\vartheta\in\Theta_d$.
Define
\begin{align}
\mathcal L_{V,\kappa}(d,\vartheta)
=
\mathcal L(\vartheta)
+
\kappa\frac{d(p+1)\log n}{n_V}.
\label{equ:gnet_pm_penalized_validation_loss}
\end{align}
Let $\hat d_L\in\mathcal M_{n_B}$ and $\hat\vartheta\in\Theta_{\hat d_L}$ denote the estimated width and parameters, and define
\begin{align}
\mathcal R_n^{\rm opt}(f_0)
=
\mathbb E_{\mathcal D,\rm alg}
\left[
\mathcal L_{V,\kappa}(\hat d_L,\hat\vartheta)
-
\inf_{\substack{d\in\mathcal M_{n_B}\\\vartheta\in\Theta_d}}
\mathcal L_{V,\kappa}(d,\vartheta)
\right].
\label{equ:gnet_pm_adaptive_optimization_gap}
\end{align}
Then, for sufficiently large $n$,
\begin{align}
\sup_{f_0\in\mathcal W_{p}^{s}}
\mathbb E_{\mathcal D,\rm alg}
\|\hat f_{\hat\vartheta}^{B}-f_0\|_P^2
\lesssim
n_B^{-\frac{2s}{2s+p}}
+
\sup_{f_0\in\mathcal W_{p}^{s}}
\mathcal R_n^{\rm opt}(f_0).
\label{equ:gnet_pm_algorithmic_generic_rate}
\end{align}
\end{theorem}
The leading statistical term $n_B^{-2s/(2s+p)}$ matches the standard $p$-dimensional Sobolev regression rate without logarithmic inflation. In contrast, related convergence bounds for FNNs with ReLU activation can contain additional logarithmic factors \cite{schmidt2020nonparametric}. If the optimization gap is of this order or smaller, the penalized model selection of neuron counts in Theorem \ref{theorem:gnet_pm_generic_rate} attains this rate. 

The theoretical results use Gaussian observational noise, constrained parameters, and model selection to obtain risk bounds. In practice, training GNet may not  impose these constraints or perform the same model-selection procedure due to the computational cost. Extending the rates to weaker noise assumptions and  the implemented optimization procedures is worth further study.

\section{Numerical studies}
\label{sec:numerical}

We consider three groups of numerical studies: three simulated functions at eight sample sizes in Section \ref{subsec:simulation}, five response variables from the UCI Machine Learning Repository \cite{kelly2023uci} in Section \ref{subsec:uci}, and prediction in classical density functional theory (cDFT) from high-dimensional unsmoothed functional inputs in Section \ref{subsec:cdft}.  We compare methods from two broad categories. 

The first class of approaches are GP, deep GP and approximate GP approaches. 
For the exact GP, denoted by GP, we use the {\tt rgasp} function in the {\sf RobustGaSP R} package, with a product Mat{\'e}rn kernel in (\ref{equ:matern_5_2}) and robust marginal posterior mode estimation \cite{gu2018robustgasp}. We estimate a nugget parameter by letting ({\tt nugget.est=T}) to accommodate observational noise and numerical or model discrepancy. 
We also include a two-layer deep GP  \cite{damianou2013deep}, implemented with {\sf R torch}, that maps the $p$ inputs to 10 independent latent GPs and then to one output GP. Each GP has 128 learned inducing points and a product Gaussian kernel with one range parameter per input variable. We optimize the variational evidence lower bound with Adam \cite{kingma2014adam} using learning rate $0.01$, minibatches of at most 1,000 observations, and 100 Monte Carlo samples through the latent layer. 
We further compare Vecchia \cite{GPGP_RPackage} and scaled Vecchia (SVecchia) \cite{katzfuss2022scaled} approximations of GPs. To improve accuracy, Vecchia uses conditioning sizes 60 and 120 in two training optimizations and 200 for prediction, while SVecchia uses 120 for training and 200 for prediction. Both settings are much larger than their package defaults to achieve higher predictive accuracy.

The second class of models includes FNNs, GNets and KANs. A $5\%$  subset of the training data is held out for selecting the best model, and parameters are estimated from the remaining fitting subset. FNN-R has three hidden layers of 512 softplus neurons, and FNN-S has two hidden layers of 30 softplus neurons, implemented with {\sf keras3 R} \cite{keras3}. For both, the activation is softplus, $\psi(x)=\log(1+e^x)$, which was more accurate than ReLU in these experiments. Their batch size is 256 and the learning rates are $0.001$ for FNN-R and $0.02$ for FNN-S, except that FNN-S uses $0.005$ in Section \ref{subsec:cdft}. GNet-1L has one layer with 10 neurons, and GNet-2L has two layers with 30 neurons per layer. Both use the Mat{\'e}rn kernel in (\ref{equ:matern_5_2}), Adam learning rate $0.02$, and model-building and validation batch sizes $n_B=\min(1000,\lfloor n/2 \rfloor)$ and $n_V=\min(1000,n-n_B)$, respectively. Five percent of the training data is held out  to  select the best epoch for approaches using Adam in optimization, and is used for UQ in GNets. KAN-1L and KAN-2L use the corresponding architectures as GNet-1L and GNet-2L, respectively, and the same learning rate and batch size as GNet. For FNN, KAN, and GNet, training uses at least 100 epochs and 2,000 Adam updates.

The compared methods cover strong GP and neural-network baselines rather than all available predictors. In  \cite{rumsey2025all}, the exact GP in \cite{gu2018robustgasp} and SVecchia in \cite{katzfuss2022scaled} are among the strongest methods at smaller and larger sample sizes, respectively, across 29 emulators, 60 synthetic test functions, and 40 real-world computer models. Both approaches are included for comparison here.

For predictive accuracy, we record the out-of-sample normalized RMSE (NRMSE)  
\begin{align}
\mbox{NRMSE}&=\left(\frac{\sum^{n^*}_{i=1} (\hat y(\mathbold x^*_i)-  y_{signal}(\mathbold x^*_i))^2}{\sum^{n^*}_{i=1} ( y_{signal}(\mathbold x^*_i)-  \bar y)^2}\right)^{\frac{1}{2}},\label{equ:nrmse}
\end{align}
where $y_{signal}(\mathbold x_i^*)$ is the latent function in Section \ref{subsec:simulation} and the observed response in Sections \ref{subsec:uci}--\ref{subsec:cdft}, while $\bar y$ is the mean of the training responses. For GP and GNet, we additionally report the empirical coverage $P_y(95\%)$, defined as the fraction of held-out responses contained in the $95\%$ predictive intervals, and the normalized average interval length $L_y(95\%)$, defined as the average interval width divided by the held-out response range $y^*_{max}-y^*_{min}$. Their formulas are given in Section S9.1 
in the Supplementary Material. Coverage closer to $95\%$ with shorter intervals is preferred, although these two criteria must be interpreted jointly. We report UQ only for GP and GNet because predictive uncertainty was not implemented for the remaining benchmarks.

\begin{figure}[t]
    \centering
    \includegraphics[scale=0.6]{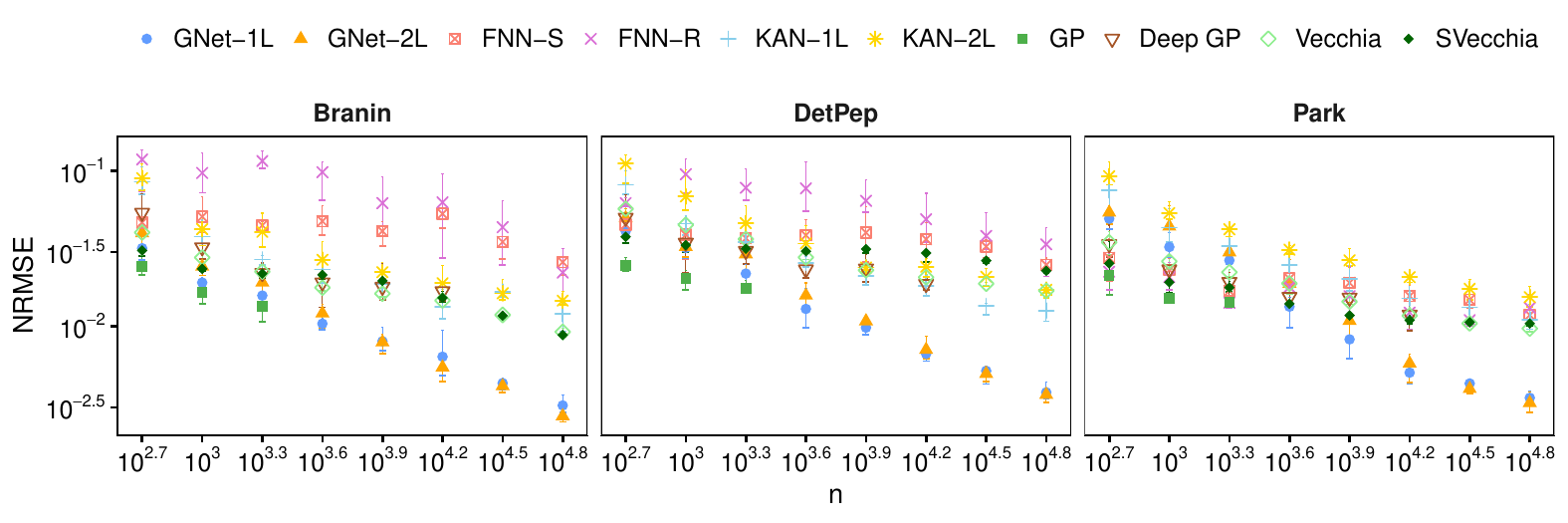}  \\
    \vspace{-.1in}
        \includegraphics[scale=0.57]{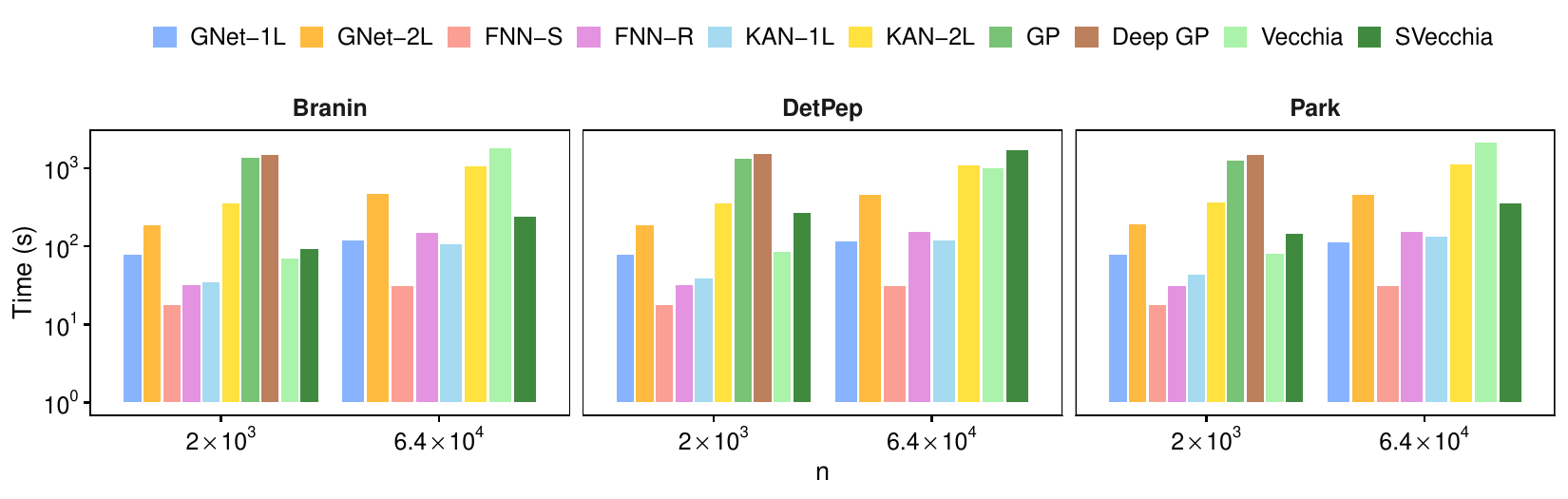}
    \caption{Comparison of NRMSE (first row) and computational time (second row) at different sample sizes on log-log scales. The arrows show the range of NRMSE across five repetitions. The exact GP is evaluated up to $n=2,000$ and the deep GP up to $n=16,000$ because of their computational cost.  }
    \label{fig:nrmse_time_simul}
\end{figure}

We record the computational time for all methods. For GP and the two GNets, the reported time includes prediction and UQ, while for the other methods it includes prediction but not UQ. Experiments use a macOS Ventura iMac with an 8-core 3.60 GHz Intel i9 processor and 32 GB RAM.  GNet, KAN and deep GP do not use parallel computation, whereas other compared approaches use the configured CPU parallel computation from their packages.  
The timing results should therefore be interpreted as implementation-level comparisons.

\subsection{Simulated experiments}
\label{subsec:simulation}

We study  Branin \cite{morris1993bayesian}, Dette--Pepelyshev curved (DetPep) \cite{dette2010generalized}, and Park \cite{park1991tuning} functions, with two to four input variables. Their expressions are given in Section S9.2 
of the Supplementary Material. Data are generated by $y(\mathbold x)=y_{signal}(\mathbold x)+\varepsilon$, where $\varepsilon$ is Gaussian noise with variance $0.01\,\mathrm{Var}\{y_{signal}(\mathbold x)\}$. We use eight training sizes, $n=500,1,000,\ldots,64,000$, uniformly sampled inputs, and $n^*=10,000$ uniformly sampled test inputs. Each experiment is repeated five times.

The NRMSEs of all compared approaches are plotted in the first row of Figure \ref{fig:nrmse_time_simul} and the second row shows the computational time at $n=2,000$ and $n=64,000$. The exact GP  gives the smallest NRMSE for these smooth low-dimensional functions from $n=500$ to $n=2,000$ (Figure \ref{fig:nrmse_time_simul}), but its $\mathcal O(n^3)$ direct computation limits larger runs. The Deep GP is evaluated through $n=16,000$. Vecchia and SVecchia reduce GP cost at larger $n$, with some loss of predictive accuracy and the larger conditioning sets used here also increase their computational cost.

From $n=8,000$ to $n=64,000$, GNet-1L and GNet-2L achieve the smallest NRMSE among the evaluated methods, shown in the first row of Figure \ref{fig:nrmse_time_simul}. At the two largest sample sizes, their average NRMSE is typically less than half that of the best non-GNet approach across the three examples.  The GNet errors also decrease approximately linearly on the log-log scale over the studied range while the architectures remain fixed. The empirical pattern at this finite range of observations is worth further investigation, because theoretical analyses of FNN regression often allow network size to increase with sample size \cite{kohler2021rate,schmidt2020nonparametric}.

Both GNet models also have relatively low cost compared with the approximate GP and neural-network variants shown in the second row of Figure \ref{fig:nrmse_time_simul}, 
reflecting the acceleration by the JIKF algorithm.  
 The reported GNet cost includes computing predictive intervals, while the other approaches, except for the exact GP, do not include predictive uncertainty quantification. Furthermore, Figure S2
 in the Supplementary Material reports predictive interval coverage and normalized interval length. Coverage is near the nominal $95\%$ level in most scenarios, while interval lengths level off as $n$ grows because the outcomes contain irreducible noise.

\begin{figure}[t]
   \centering
\includegraphics[scale=0.5]{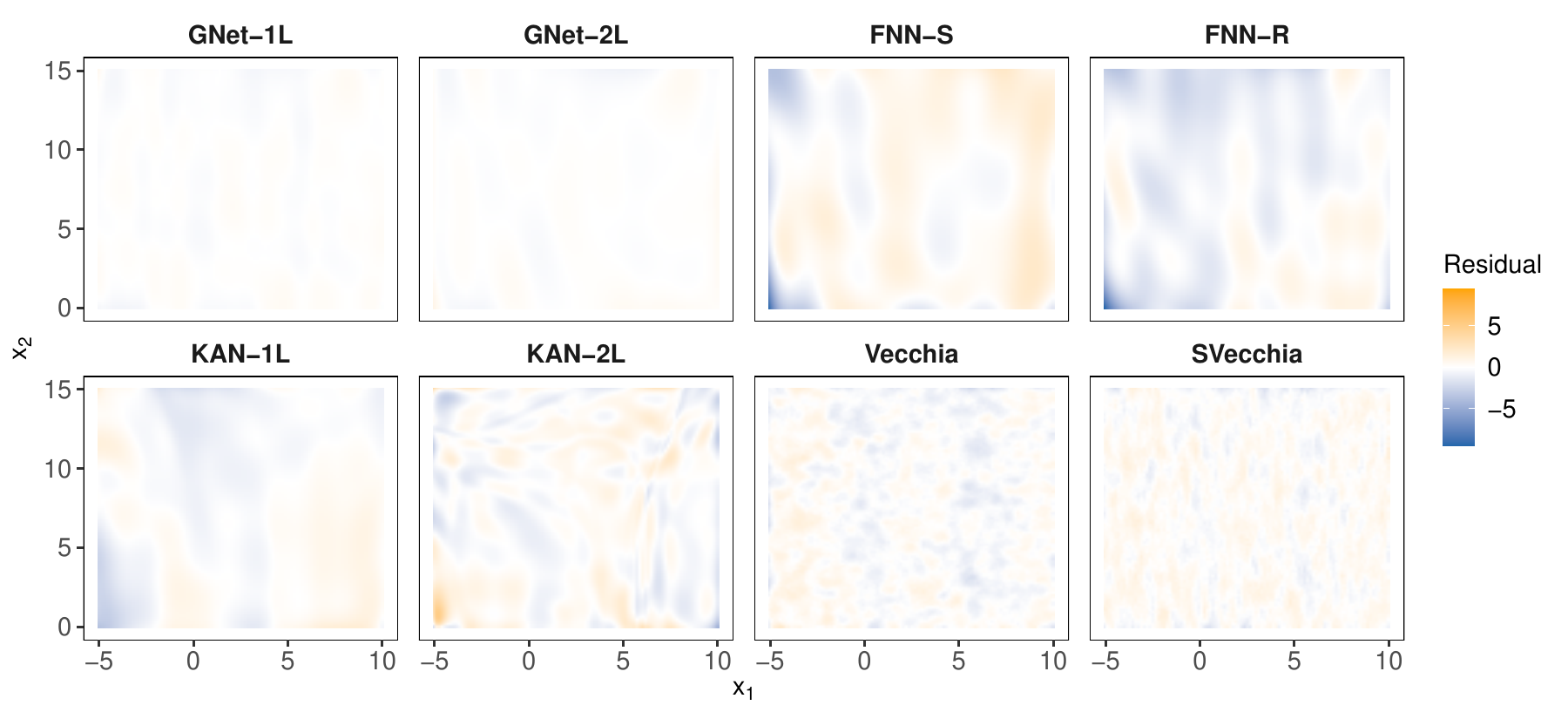}
    \caption{Prediction residuals for one Branin experiment at $n=64,000$. The standard deviation of the latent signal is approximately 52.21.}
    \label{fig:residual_branin}
\end{figure}

 Figure \ref{fig:residual_branin} shows the residuals between the predictions and the truth in one Branin experiment at the largest sample size. The residuals of all methods are  small compared with the standard deviation of the signal, while the GNet residuals are visually smaller than those of the other approaches.
 The estimated final-layer neuron functions of GNet-1L and GNet-2L for one experiment in each testbed are plotted in Figure S3 
in the Supplementary Material. These functions illustrate the flexibility of the learned nonparametric activations in compact GNet models.

\subsection{Nonparametric regression for  real-world data}
\label{subsec:uci}

\begin{figure}[t]
\centering
    \centering
    \includegraphics[scale=0.6]{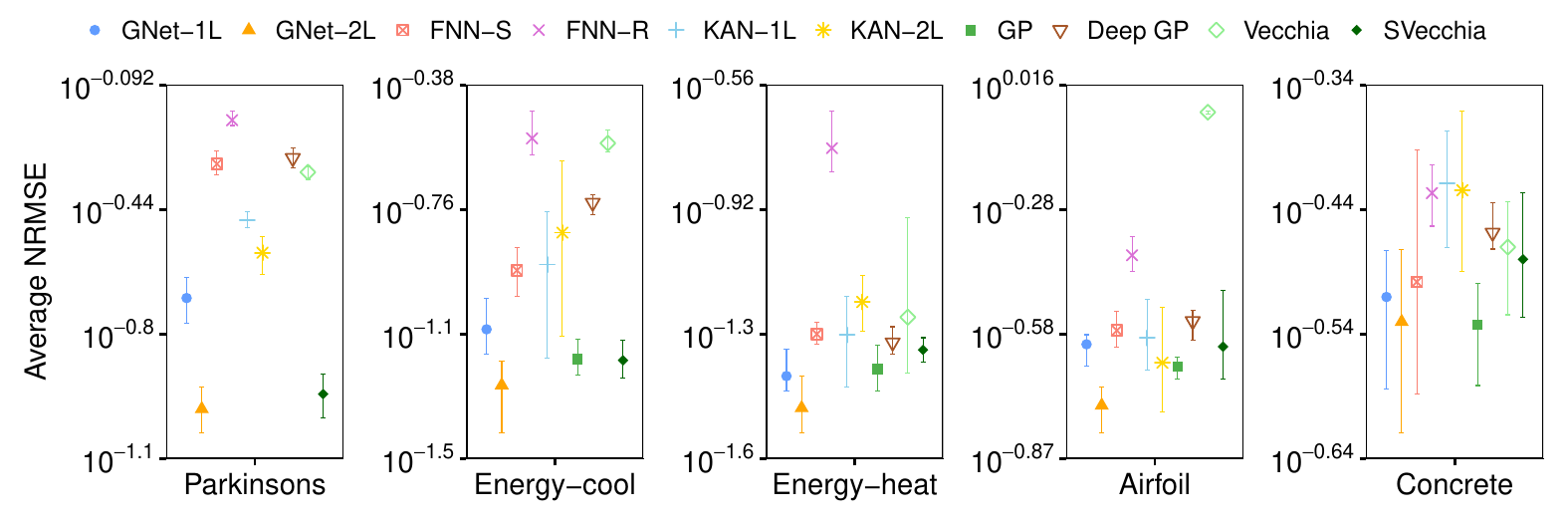}
    \caption{The average NRMSE for five response variables from four UCI data sets. The total sample size $n_{all}$ and input dimension $p$ are: Concrete ($n_{all}=1030$, $p=8$), Parkinsons ($n_{all}=5875$, $p=19$), Energy-heat and Energy-cool ($n_{all}=768$, $p=8$), and Airfoil ($n_{all}=1503$, $p=5$).  }
    \label{fig:uci_nrmse}
\end{figure}

We consider Parkinson's telemonitoring (Parkinsons) \cite{tsanas2009accurate}, energy efficiency with heating and cooling loads (Energy-heat and Energy-cool) \cite{tsanas2012accurate}, airfoil self-noise (Airfoil) \cite{airfoil_self-noise_291}, and concrete compressive strength (Concrete) \cite{yeh1998modeling}. For each dataset, we use five-fold cross-validation, with $80\%$ of the data for model training and the remaining $20\%$ for testing in each fold. These scenarios correspond to training sample sizes of around 500 to 5,000 to compare predictive performance.

Figure \ref{fig:uci_nrmse} shows that GNet-2L has the smallest average NRMSE in the first four experiments, while the exact GP is slightly more accurate for the Concrete dataset. For the Parkinsons dataset, the  split evaluates prediction of held-out recordings instead of predicting records from new patients. SVecchia has NRMSE close to GNet-2L, but its average computational time is about four times larger, as shown in Figure S4 
in the Supplementary Material. 
Thus GNet-2L remains competitive at these small-to-moderate sample sizes despite using the same architecture and optimization setting as in the simulation in Section \ref{subsec:simulation}.

The average normalized 95\% predictive interval lengths and empirical coverages for GP, GNet-1L, and GNet-2L are plotted in Figure S5 
in the Supplementary Material. The exact GP results are not available for the Parkinsons dataset due to its high computational cost. 
For all three approaches, coverage of the holdout test data is generally close to the nominal level. Some GNet coverages fall slightly below the nominal level for the Energy responses. The normalized predictive interval lengths are relatively short, indicating informative uncertainty quantification.

\subsection{Classical density functional theory  with high-dimensional inputs}
\label{subsec:cdft}
\begin{figure}[t]
    \centering
    \includegraphics[scale=0.48]{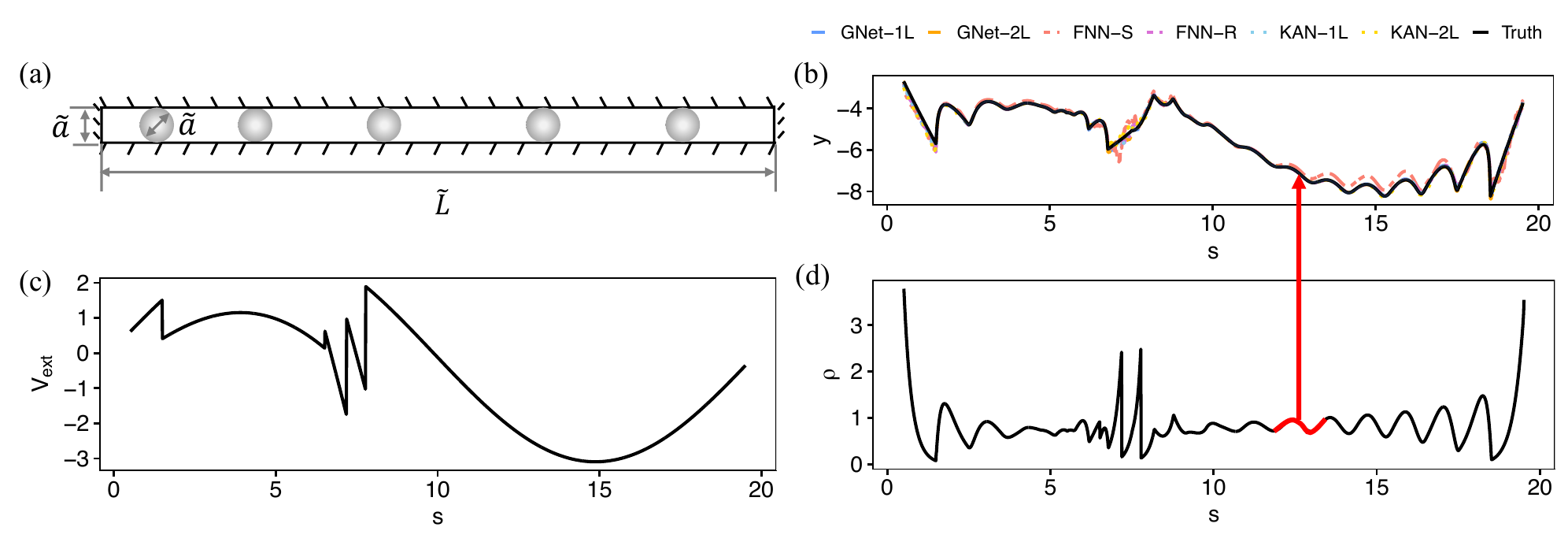}
    \caption{(a) One-dimensional hard-rod system with particle length $\tilde a$ and domain length $\tilde L$. (b) Truth and predictions of the one-body direct correlation function, and the curves almost overlap. The nearly unnoticeable blue shaded area is the $95\%$ predictive interval from GNet-1L. (c) One sampled external potential $V_{\rm{ext}}(s)$. (d) The corresponding particle density. The red segment  highlights the 201 local particle density values  used to predict one value  in panel (b).  }
    \label{fig:cdft_illustration}
\end{figure}

Finally, we predict the one-body direct correlation function in classical density functional theory (cDFT), an important quantity for modeling inhomogeneous fluids \cite{sammuller2023neural}. We consider the one-dimensional hard-rod system, whose grand potential is
\begin{align}
    \Omega[\rho] =   f_{\rm id}[\rho]+ f_{\rm ex}[\rho]+\int^{\infty}_{-\infty} \rho(s)[V_{{\rm ext}}(s)-\mu_{\rm chem}]  ds ,
    \label{equ:Omega}
\end{align}
where $\rho(s)$ is the particle density, $V_{\rm ext}$ is the external potential, $\mu_{\rm chem}$ is the chemical potential, and $f_{\rm id}$ and $f_{\rm ex}$ are the ideal-gas and excess free-energy functionals. Following \cite{fang2022reliable}, the inverse thermal energy and thermal wavelength are set to one, $\beta_{\rm th}=\Lambda=1$. The ideal-gas functional satisfies $\beta_{\rm th}f_{\rm id}[\rho]=\int \rho(s)\{\log[\rho(s)\Lambda^3]-1\}ds$. For hard rods of length $\tilde a$, the excess free-energy functional follows $\beta_{\rm th}f_{\rm ex}[\rho]=-\int \rho(s)\log\{1-\int_{s-\tilde a}^{s}\rho(s')ds'\}ds$, where $\tilde a$ is the particle length. The integral satisfies $\int_{s-\tilde a}^s\rho(s')ds'\leq 1$ because any interval of length $\tilde a$ can contain at most one particle.

A schematic overview of the system is plotted in panel (a) in Figure \ref{fig:cdft_illustration}. 
The equilibrium density profile minimizes $\Omega$ and thus satisfies the Euler-Lagrange (E-L) equation \cite{fang2022reliable}: 
\begin{align}
    \rho(s) = \exp \left\{ \mu_{\rm chem} - V_{\rm ext}(s) - \frac{\delta f_{\rm ex}[\rho]}{\delta\rho}(s)\right\},
    \label{equ:rho_EL}
\end{align}
where ${\delta f_{\rm ex}[\rho]}/{\delta\rho}$ denotes the functional derivative of excess free-energy $f_{\rm ex}$ with respect to particle density $\rho$. In a 1D hard-rod system, the target one-body direct correlation function $y[\rho](s)=-\beta_{\rm th}\delta f_{\rm ex}[\rho]/\delta\rho(s)$, with its expression provided in Section S9.4 
of the Supplementary Material, depends only on the density over $[s-\tilde a,s+\tilde a]$ for hard rods. This reduces the relevant input from the full system to a local density window of length twice the particle size,  $2\tilde a$ \cite{sammuller2023neural}.

Our goal is to predict the one-body direct correlation function $y[\rho](s)$ at all $s$ from the local particle density over length $2\tilde a$. Following \cite{sammuller2023neural}, we sample $\mu_{\rm chem}\sim\mathcal U(-5,10)$, with $\mathcal U$ denoting the uniform distribution, and generate a broad class of external potentials
\begin{equation}
  V_{\rm{ext}}(s)
  =\sum_{t'=1}^{4} A_{t'}\sin\left(\frac{2\pi st'}{\tilde L}+\phi_{t'}\right)
   +\sum_{t=1}^{m_{lin}} v^{\rm lin}_{t}(s),
  \label{equ:cdft_generated_vext}
\end{equation}
for $\tilde a/2 \leq s \leq \tilde L-\tilde a/2$, where $A_{t'}\sim\mathcal N(0,2.5)$, $\phi_{t'}\sim\mathcal U(0,2\pi)$, and $m_{lin}$ is uniform on $\{1,\ldots,5\}$, and $V_{\rm ext}=+\infty$ elsewhere. Each piecewise-linear component is
$ v_t^{\rm lin}(s)=v_{t,1}+(v_{t,2}-v_{t,1})(s-s_{t,1})/(s_{t,2}-s_{t,1})$ for $s_{t,1}<s<s_{t,2}$ and zero elsewhere, where $v_{t,1},v_{t,2}\sim\mathcal N(0,4)$ and $s_{t,1},s_{t,2}\sim\mathcal U(0,\tilde L)$, and the two locations are swapped when necessary so that $s_{t,1}<s_{t,2}$. We solve (\ref{equ:rho_EL}) numerically by the Picard algorithm, and the details are given in Section S9.4
of the Supplementary Material.  We set $\tilde L=20$, $\tilde a=1$, and grid spacing $\Delta_s=0.01$. At each location, the predictor is the $p=201$ vector $\mathbold x(s)=[\rho(s-\tilde a),\rho(s-\tilde a+\Delta_s),\ldots,\rho(s+\tilde a)]^T$. 
       
       Figure \ref{fig:cdft_illustration}(c)--(d) illustrates a sampled external potential containing sharp changes and the resulting unsmoothed density. The red segment highlights the 201 local density values used as input to predict one value of the one-body direct correlation function. These high-dimensional, unsmoothed functional inputs pose a challenge for prediction. Furthermore, the particle density is positive, $\rho(s) \in \mathbb R^+$, while the one-body direct correlation function is a real-valued quantity, $y[\rho](s)\in \mathbb R$.
       These physical quantities are distinct from probability density and correlation.

We generate 100 density profiles with the last $50$ profiles held out for testing, and use $10, 20, $\ldots$, 50$ of the remaining profiles for training. Because particles can appear only in $[\tilde a/2, \tilde L-\tilde a/2]$, each profile contributes 1,901 interior locations, giving training sizes $n=19,010,38,020,\ldots,95,050$ and $n^*=95,050$ observations for testing. The input dimension is $p=201$.

\begin{figure}[t]
    \centering
        \begin{tabular}{cc}
        \begin{overpic}[scale=0.5,    trim={0 -25 0 0}]{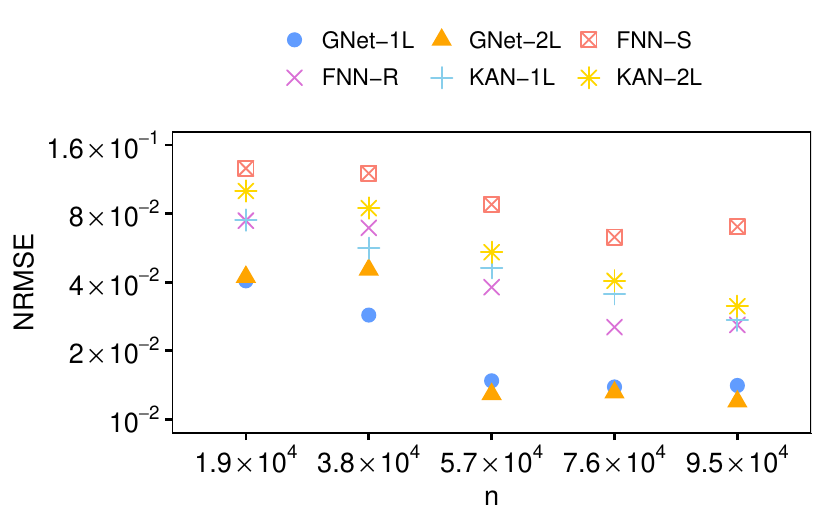}
            \put(0.5,65)
            {{(a)}} 
        \end{overpic} &
        \begin{overpic}[scale=0.5,  trim={0 -25 0 0}]{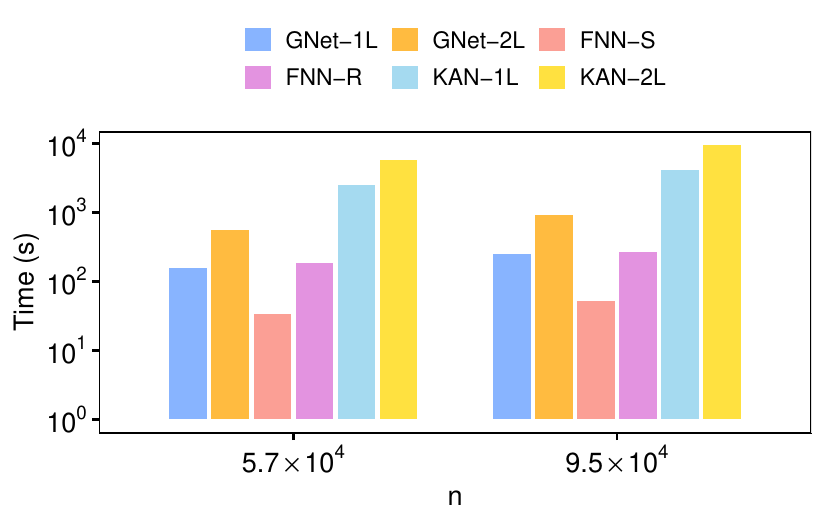}
            \put(0.5,65){{(b)}}
        \end{overpic} \vspace{-.25in}\\
             \end{tabular}
        \vspace{-.1in}
    \caption{(a) NRMSE across training sample sizes. (b) Computational time at two sample sizes. 
     }
    \label{fig:cdft_nrmse_time}
\end{figure}

We found that the exact GP,  Vecchia, and SVecchia from their standard packages, as well as the implemented Deep GP, were not computationally competitive within the budgets considered for these high-dimensional settings. For Vecchia and SVecchia, we observed substantial neighbor-search costs when both $n$ and $p$ were large, and instability in kernel-parameter optimization for these highly similar functional inputs. Thus, we do not include them for comparison.

 Figure \ref{fig:cdft_nrmse_time} compares six approaches. 
GNet-1L and GNet-2L have the smallest NRMSE among the evaluated approaches, including FNN-R, which uses the three-512-hidden-layer softplus architecture in \cite{sammuller2023neural}. At $n=95,050$, the NRMSEs of GNet-1L and GNet-2L are approximately $0.013$ and $0.012$, while all other evaluated methods exceed $0.030$. GNet-1L is more accurate at smaller training sizes. At $n=95,050$ and $p=201$, fitting GNet-1L, predicting the test set, and computing UQ take about five minutes on the desktop computer.

Predictions for three held-out density profiles are shown in Figure S6 
of the Supplementary Material. Figure S7 
in the Supplementary Material shows that empirical coverage moves toward $95\%$ as the training size increases. Predictive uncertainty can be used to decide whether a test case is sufficiently reliable for an error-control procedure \cite{fang2022reliable}.

\section{Concluding remarks}
\label{sec:conclusion}

We develop GNets and fast algorithms for scalable and efficient predictions. We demonstrate in both theoretical and numerical studies its high functional flexibility without requiring model and sample sizes to grow as rapidly as in some neural-network models.  GNet also opens doors for a wide range of research directions.  GNet can be extended to other neural-network architectures for categorical or vectorized outputs. Reliable uncertainty quantification with little loss of predictive accuracy or computational efficiency remains a future study. Alternative optimization methods may further improve training, particularly because GNets can use substantially fewer parameters than conventional neural networks. Furthermore, the current risk bound theory focuses on one-layer GNets, and extending it to multilayer architectures with more general assumptions is an important direction.

\section*{Acknowledgement}
 We acknowledge the support from the National Science Foundation under Award No. OAC-2411043. 
\bibliographystyle{plain}
\bibliography{References_chronical_2026}

\end{document}